\documentclass[fleqn,usenatbib]{mnras}

\usepackage{newtxtext,newtxmath}

\usepackage[T1]{fontenc}

\DeclareRobustCommand{\VAN}[3]{#2}
\let\VANthebibliography\thebibliography
\def\thebibliography{\DeclareRobustCommand{\VAN}[3]{##3}\VANthebibliography}

\makeatletter
\def\endfigure{\end@float}
\makeatother

\makeatletter
\def\endtable{\end@float}
\makeatother

\usepackage{graphicx}
\usepackage{amsmath}

\usepackage{amssymb}
\usepackage{multicol}       
\usepackage{booktabs}
\usepackage{newtxtext,newtxmath}
\usepackage{nicematrix}
\usepackage{caption}
\usepackage{enumitem}
\usepackage[utf8]{inputenc}
\usepackage{makecell}
\usepackage{threeparttable}
\usepackage{tabularray}
\usepackage{cleveref}
\usepackage{array}
\usepackage{float}
\usepackage{subfigure}
\usepackage{hyperref}
\usepackage{tabularx}
\usepackage{longtable}

\title[Exploring Globular Cluster Dynamics with CVs]{Probing Intracluster Dynamics and Evolution of Globular Clusters through Cataclysmic Variable Populations}

\author[K. Oh et al.]{
Kwangmin Oh,$^{1,2}$
Jongsuk Hong,$^{3}$
C. Y. Hui,$^{4}$\thanks{E-mail: huichungyue@gmail.com, cyhui@cnu.ac.kr}
Sangin Kim$^{2}$
and Mirek Giersz$^{5}$
\\
$^{1}$Department of Physics and Astronomy, Michigan State University, East Lansing, MI 48824, USA\\
$^{2}$Department of Space Science and Geology, Chungnam National University, Daejeon 34134, Republic of Korea\\
$^{3}$Korea Astronomy and Space Science Institute, Daejeon 34055, Republic of Korea\\
$^{4}$Department of Astronomy and Space Science, Chungnam National University, Daejeon 34134, Republic of Korea\\
$^{5}$Nicolaus Copernicus Astronomical Centre, Polish Academy of Sciences, ul. Bartycka 18, PL-00-716 Warsaw, Poland\\
}

\date{Accepted XXX. Received YYY; in original form ZZZ}

\pubyear{2024}

\begin{document}
\label{firstpage}
\pagerange{\pageref{firstpage}--\pageref{lastpage}}
\maketitle

\begin{abstract}
Dynamical interactions in globular clusters (GCs) significantly impact the formation and evolution of binary sources, including cataclysmic variables (CVs). This study investigates the connection between dynamical states of GCs and X-ray luminosity ($L_{x}$) distributions of CV populations through both simulations and actual observations. Utilizing a Monte Carlo simulation tool, MOCCA, we categorize the simulated GCs into three different evolutionary stages which are referred to as Classes I/II/III. Significant differences are found in the $L_{x}$ distributions of the CVs among these three Classes. In observational aspects, we have analyzed 179 CV candidates in 18 GCs observed by the {\it Chandra} X-ray Observatory.  By dividing these GCs into three Families of different dynamical ages, namely Families I/II/III, the $L_{x}$ distributions of the CV candidates also show significant differences among these three Families. Both simulations and observational results suggest that CVs in more dynamically evolved clusters (Class/Family III) exhibit brighter X-ray emission. This highlights the influence of the dynamical status of a GC on the properties of its hosted compact binaries. Similar to blue stragglers, CV populations can serve as tracers of a GC's dynamical history. Our findings provide insights for understanding the interplay between intracluster dynamics and the evolution of compact binaries in GCs.
\end{abstract}


\section{Introduction}
The complex interplay between the dynamical formation of binary sources, including cataclysmic variables (CVs) and the evolution of globular clusters (GCs) has been an intriguing topic for numerous studies over the years \citep[e.g.][]{Ivanova_2006, Hong_2017}. CVs are binaries comprised of a white dwarf accretor and a donor star which are typically low-mass main-sequence (MS) stars. Mass transfer occurs when the donor star fills its Roche lobe which leads to the accretion of material on the surface of the white dwarf and gives rise to the X-ray emission. For the CVs hosted by GCs, their formation and evolution can be complicated by frequent stellar encounters. The intracluster dynamics such as tidal capture and exchange interaction \citep[]{Hut_1992, Pooley_Hut_2006,belloni1,belloni2,belloni3,belloni4} can strongly affect the binary properties of CVs in GCs (e.g., mass ratio, orbital separation). This can possibly result in emission properties that differ from those of binaries in an environment without such a dynamical process (e.g., Galactic field). This notion is supported by the recent discovery that some properties of millisecond pulsars (MSPs) in GCs (e.g., X-ray conversion efficiency, rotational period distribution) are significantly different from their counterparts in the Galactic field \citep{Lee_2023}, providing evidence for the influence of intracluster dynamics on the evolution of their progenitors (e.g., low-mass X-ray binaries (LMXBs)).

While intracluster dynamics can affect the evolution of compact binaries, the presence of such exotic systems in GCs can also influence cluster evolution. Elastic scattering between binary systems and single stars or other binaries can prevent or reverse core collapse by transferring gravitational potential energy to the kinetic energy of stars in the core \citep{Hut_1992}. If the initial binding energy of a primordial binary is larger than the average kinetic energy of the surrounding stars in a GC, the encounters between them can lead to orbital shrinkage with the orbital binding energy transferred to the others \citep{heggie_1975}, which is commonly referred as binary burning which can drive the long-term evolution of a GC.

In GCs, stellar interaction can lead to energy equipartition among stars/binaries. However, achieving complete energy equipartition is unlikely in realistic initial mass functions for GC, as suggested by \cite{Trenti_2013}. Nevertheless, this tendency leads to the inward movement of massive systems towards the centre which is commonly referred to as dynamical mass segregation. 

For the binaries with their orbits tight enough, they can be considered single objects with masses equal to the sum of their stellar components. Consequently, they also undergo dynamical mass segregation. Since the central regions have higher stellar density and a larger fraction of high-mass stars, binaries that have segregated towards the centre have a larger likelihood of evolving into exotic binary systems, such as MSPs, CVs, and LMXBs, through various dynamical processes such as orbital hardening and the replacement of members with higher masses.

In this context, the role of black holes becomes particularly significant. Mass segregation within stellar systems predominantly impacts the most massive stars or star groups, which include black holes that may be numerous in GCs due to processes like mass fallback \citep{Belczynski_2002}. When a significant black hole (BH) subsystem is present, such subsystems can undergo rapid core collapse. This phenomenon is distinct from the segregation and collapse processes affecting ordinary stellar systems, including progenitors of CVs. In fact, for these ordinary stars to undergo mass segregation and core collapse, the BH subsystems typically need to be expelled or sufficiently reduced \citep{breen_2013}. Additionally, the gravitational influence of an intermediate-mass black hole can significantly modify this dynamic, often impeding the segregation of binary systems like CVs \citep{Hong_2020}.

According to the photometric concentration, GCs are commonly classified as core-collapsed or non-core-collapsed based on their core sizes, which might correspond to the pre-core-collapse and post-core-collapse phases \citep{Harris1996.112}. However, this classification may not sufficiently reflect the dynamical state of a GC. The process of mass segregation, including binaries, in clusters begins with core contraction by two-body relaxation and continues until core collapse occurs, as described by \cite{Giersz_heggie_1996} and \cite{Hong_2013}. The investigation of exotic binaries whose formation is related to the dynamical encounters can offer a more nuanced perspective on this classification. 

According to \cite{Ferraro_2012}, GCs can be categorized into three different groups according to their radial distributions of blue stragglers (BSs), which is different from the previously mentioned binary classification algorithm. \cite{Ferraro_2012} refers to these divisions as Families I/II/III, which reflect how the BSs are segregated. Family I GCs have a flat BS radial distribution, indicating insufficient time for mass segregation. As an illustration, Family II's BS distribution shows a dip, meaning that mass segregation has only affected BSs close to the centre. Strongly segregated BSs are prominently displayed in the core region of Family III GCs, which have a centrally peaked BS distribution. 

In order to understand how the evolution of GCs can lead to trifurcation, we utilize a Monte Carlo code to explore how cluster properties vary from zero age to present. The simulation can also enable us to examine how the binary population co-evolves with the cluster. On the other hand, using the actual X-ray observations, we examine whether the properties of the X-ray binaries differ among Families I/II/III.

In this work, we focus on the quiescent X-ray luminosity of CVs. First, in comparison with the other classes of compact binaries such as MSPs, the accretion-powered X-ray emission from CVs can be computed by a simple model in our simulations. Also, from the observational aspects, CV candidates can be easily identified by their luminous and hard X-rays. Therefore we can collect an observed sample of CV-like X-ray sources to compare with the simulated results.

\section{Previous works on dynamical formations of CVs in GCs}

Numerous studies have aimed to understand the complex interplay between stellar interactions, binary evolution, and the dynamical state of GCs. 

In the high-density environments of GCs, stellar interactions can significantly influence the formation of X-ray binaries, including CVs \citep{clark_1975, Hut_1992, verbunt_2001}. A strong correlation has been observed between the number of X-ray sources and the encounter rate of GCs, indicating that dynamical interactions play a crucial role in the formation of X-ray binaries \citep{verbunt_hut_1987, Pooley_2003, Hui_2010}. On the other hand, some numerical/theoretical studies suggest that the presence of BHs can dominate the dynamical interactions in GCs and affect the overall binary population \citep{breen_2013, 2022MNRAS.509.4713W}.

Recently, a study on MSPs in GCs has provided a deeper insight into how the dynamical state of a GC can affect the binary evolution \citep{oh_2023}. Their study found that the MSPs in core-collapsed GCs generally rotate slower and are more luminous in radio. This suggests that the more intense dynamical interactions in core-collapsed GCs can disrupt the LMXBs at an earlier stage of the recycling process.

A significant contribution to our understanding of the dynamics of CV formation in GCs comes from \citet{belloni1, belloni2, belloni3, belloni4}. They noted that dynamics not only disrupt binaries and promote their hardening but also compel wider binaries, which would not form CVs in field evolution, to become CVs. They also emphasized the crucial role of the initial properties of primordial binaries in shaping the CV population in GCs. Furthermore, they highlighted the existence of a time delay between the central density (or rate of interactions) and the number of CVs due to the mass segregation of CVs ejected from the core.

\cite{RiveraSandoval:2017itj} also revealed that core-collapsed GCs tend to exhibit bimodal CV populations in their near-UV and optical luminosities. This discovery highlights the significance of comparing CV populations in different types of GCs, such as core-collapsed and non-core-collapsed clusters, in understanding the influence of dynamical interactions on the formation and evolution of CVs. By investigating the differences in CV properties among different classes of GCs, one can draw valuable insights into the processes that govern the formation, destruction, and overall properties of CVs in their different environments.

Despite the progress in this field, theoretical calculation of the properties of CV population based on dynamical status of a GC remains to be challenging because of the complex intracluster dynamics \citep{pryor_meylan_1993, fregeau_2009, Kremer_2019, belloni1,belloni2,belloni3,belloni4}. To exacerbate the problem, the factors concerning the evolution of compact binaries, such as mass transfer, common envelope, and orbital separation, the formation and destruction of CVs in GCs further the progress \citep{Paczynski_1976, Iben_livio_1993}. In this work, using a Monte Carlo code and a simple accretion model, we examine how the cluster density and the CV population evolve with time. By defining a present-day population at an epoch of 12 Gyr, we compute the X-ray luminosity distributions of CVs from our simulations which can be compared with that of the observed sample. 

\begin{figure*}
\centering
\subfigure{
\includegraphics[width=0.32\textwidth]{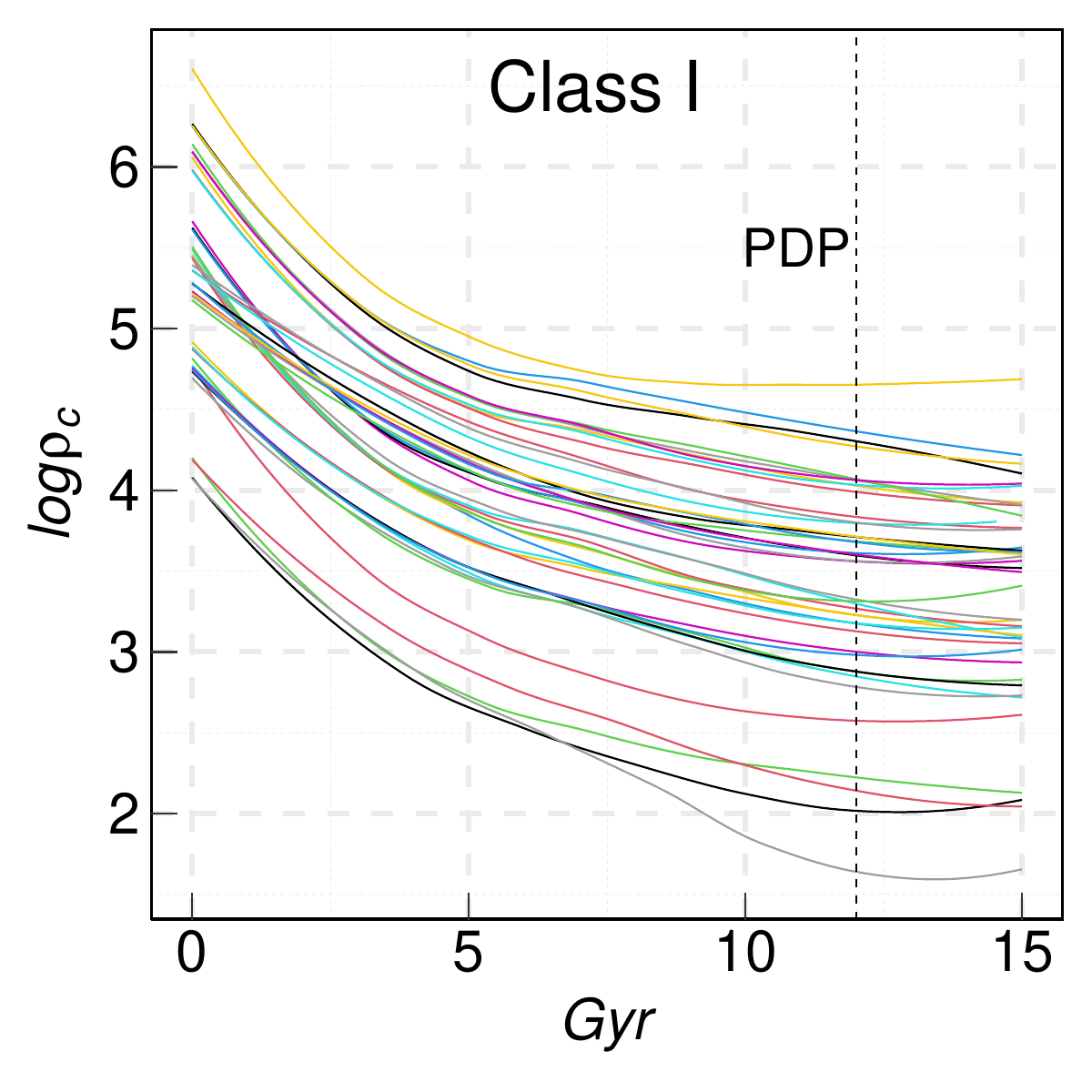}
}
\subfigure{
\includegraphics[width=0.32\textwidth]{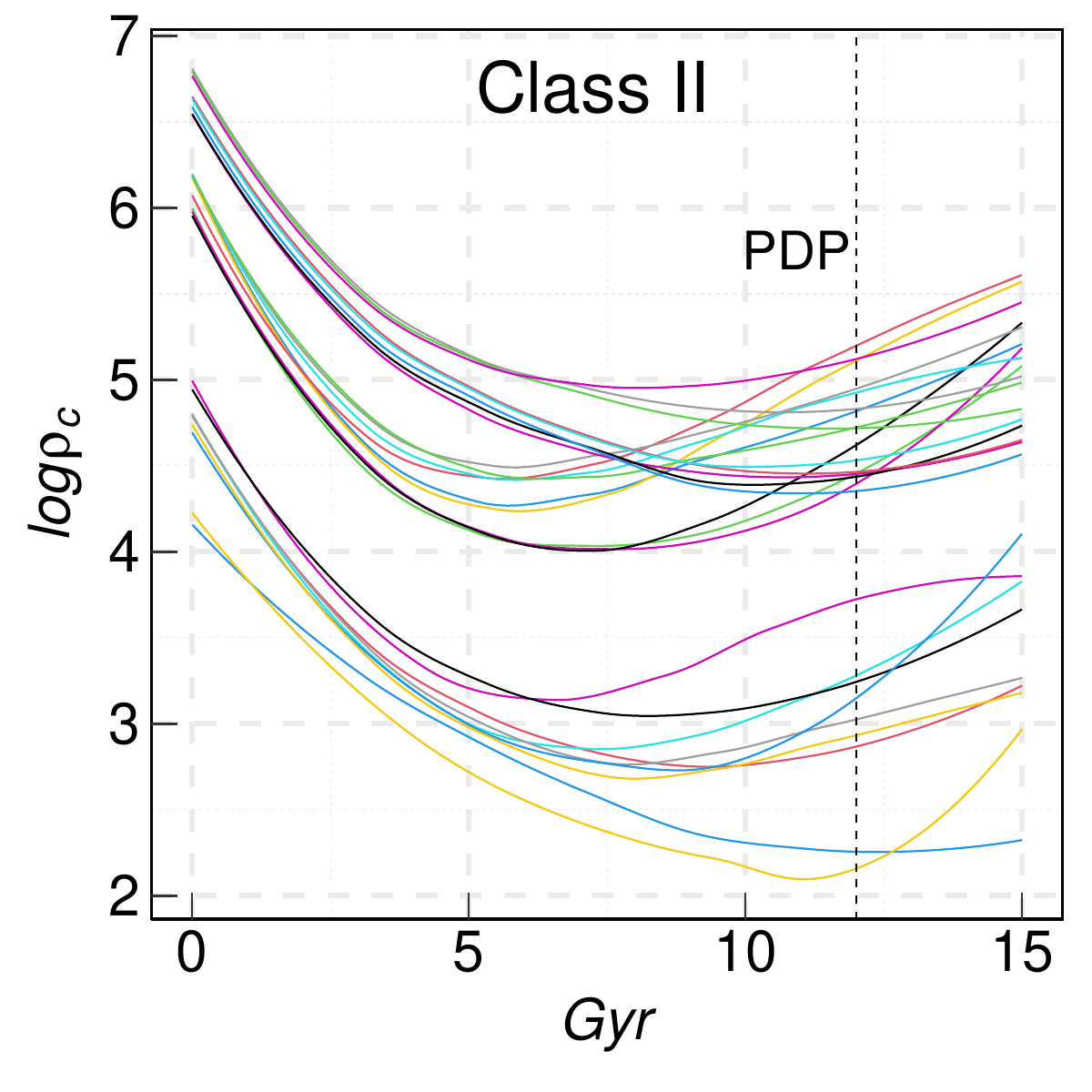}
}
\subfigure{
\includegraphics[width=0.32\textwidth]{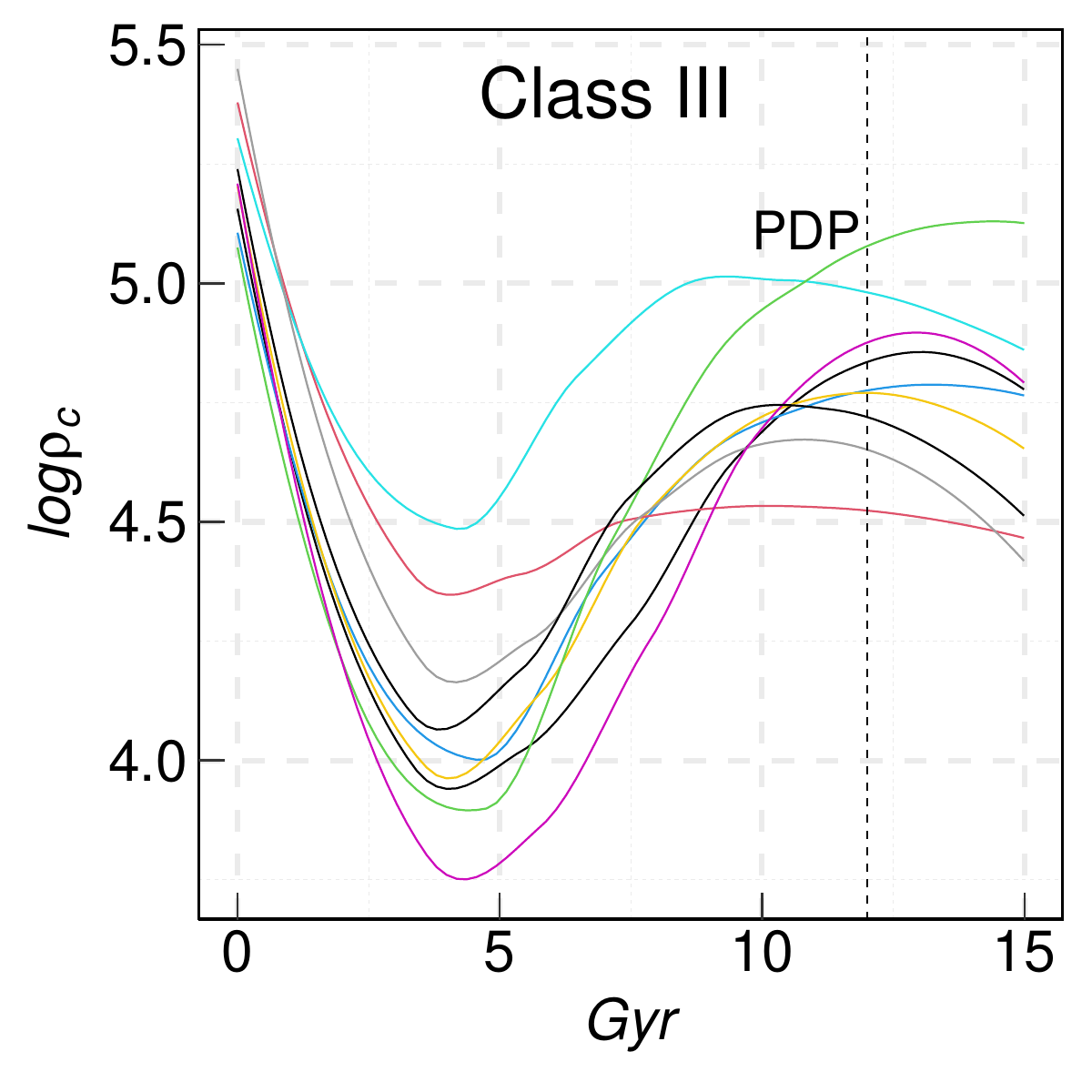}
}
\caption{Division of simulated models based on its core density, illustrating the evolution of CV populations in GCs. The three classes display the overall trends of core density over time, reflecting the different dynamical states of the clusters. The dotted line represents PDP at 12 Gyr, serving as a reference point for comparing the behavior of CV populations in relation to the dynamical history of their host clusters.}
\label{fig:model_division}
\end{figure*}

\begin{figure}
\includegraphics[width=0.9\columnwidth]{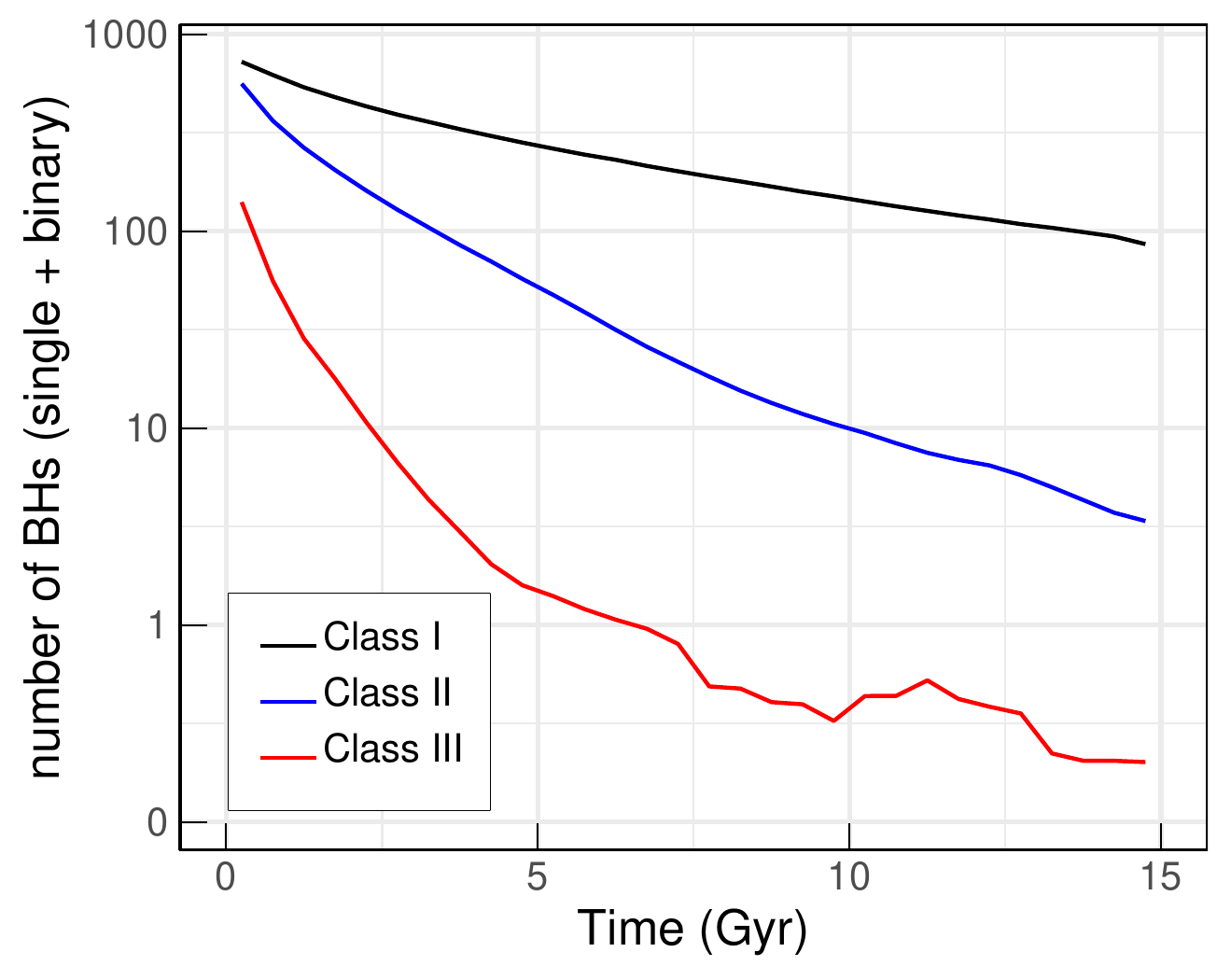}
\caption{Evolution of averaged BH numbers in MOCCA models. Class I shows the highest number of black holes, indicating their role in stabilizing the system and delaying the collapse, with Class III collapsing early in its evolution. All BHs included in binary systems are accounted for in these counts.}
\label{fig:evolution_BH}
\end{figure}

\section{Globular Cluster Simulation}
\subsection{MOCCA}

To explore the relationship between the dynamical state of GCs and the properties of their CV populations, we conducted simulations using the MOCCA (MOnte Carlo Cluster simulAtor) code. MOCCA is a robust tool for simulating the evolution of star clusters, accounting for both stellar dynamics and stellar evolution processes \citep{Giersz_2013}.

Our study employs the MOCCA code for simulating the evolution of star clusters, incorporating both stellar dynamics and stellar evolution processes. The simulations are based on a \cite{Kroupa_2001} stellar initial mass function, with stellar masses spanning from 0.1 to 100 $M_\odot$ and metallicity of Z = 0.001. A key feature of our simulations is the inclusion of primordial binaries, following the initial binary distribution as described by \cite{Kroupa_1995, Kroupa_2013}. This distribution is derived using an eigen-evolution procedure, which is a method that accounts for early dynamical interactions and stellar evolution to transform a birth population into an observed population.

In our MOCCA simulations, we made several assumptions about stellar and binary evolution processes. We employed the prescriptions for stellar winds, compact object formation, and the treatment of the tidal field from the binary stars evolution (BSE) code \citep{Hurley_2000, Hurley_2002} that is incorporated into MOCCA. Additionally, our model includes considerations of black hole kicks, which are the velocities imparted to black holes at birth due to asymmetric supernova explosions, and mass fallback, where a portion of the expelled material during the supernova falls back onto the compact object, influencing the mass and spin of black holes. These factors are critical in understanding the formation and properties of black hole subsystems in star clusters. We explored a variety of initial conditions, varying parameters such as the initial masses, half-mass radii, galactocentric distances, and primordial binary fractions of the clusters.

The initial density profile of all systems follows a \cite{king_1966} model with a central dimensionless potential, $W_0$, which is equal to 7. This choice of $W_0$ is commonly used to model GCs and serves as a representative value for our simulations. The simulations also account for the effects of a tidal cut-off, determined by the cluster mass and galactocentric distance. The ratio of the half-mass radius to the tidal radius in our simulations varies from approximately 0.005 to 0.09, reflecting a range of initial conditions.

The formation of CVs depends on the initial binary star distribution, with one primary star evolving into a white dwarf and the other staying in the main sequence. Key factors for CV formation include a significant mass difference from the start and a close enough approach to enter a common envelope phase. Typically, binaries with short periods, high eccentricities, and large mass ratios, evolve into CVs. 

We created 81 GC models encompassing a wide variety of potential configurations by varying initial conditions. The chosen initial conditions—variations in binary fraction, galactocentric distance, the initial number of particles, and the cluster's half-mass radius—were selected due to their potential impact on the dynamical state of GCs and the properties of CV populations.

The MOCCA simulations were set up as follows:
\begin{itemize}[leftmargin=*]
\item \textit{Initial number of particles:} We simulated GCs with initial particle numbers of 200K, 500K, and 1M to represent the range of GC sizes in the Milky Way.
\item \textit{Galactocentric distance:} We varied the galactocentric distances of the simulated GCs (4, 8, and 16 kpc) to examine the effects of tidal forces and the galactic potential on the clusters' dynamical evolution and the properties of their CV populations. Escaping stars were handled using the tidal boundary treatment in the MOCCA code, which accounts for both the tidal radius and the galactic potential.
\item \textit{Half-mass radius of the cluster:} Initial half-mass radii of 1, 2, and 4 pc were considered to account for the potential influence of the initial density profile on dynamical interactions and the properties of the CV populations in GCs.
\item \textit{Binary fraction:} Initial binary fractions of 10\%, 20\%, and 50\% were considered to account for the potential impact of the proportion of binary stars on dynamical interactions and CV formation in GCs.
\end{itemize}

We collected CVs at 12 Gyr, treating them as the present-day population (PDP). These CVs, comprised of WD and MS stars, were selected based on their Roche lobe overflow (RLOF)—a key feature relevant to understanding the dynamical evolution of GCs. Additionally, we traced the presence of intermediate-mass black holes (IMBHs) in our simulations. We observed IMBH formation in the densest systems with higher initial particle counts. However, it's worth noting that these IMBHs did not grow through fast scenarios, likely due to the relatively lower initial densities of our models \citep{Giersz_2015}.

To study the X-ray luminosity of each identified CV, we computed their luminosities based on the properties of their donor and WD components. Following \cite{belloni1}, we calculated the X-ray luminosity for slowly rotating WDs in the [0.5 -- 10 keV] band with the equation:

\begin{equation}
L_{\rm X} \ = \ \varepsilon~ \frac{GM_{\rm WD} \dot{M}_{\rm dQ}}{2 R_{\rm WD}} \label{XRAY}
\end{equation}

\noindent
Using the efficiency factor $\varepsilon$ (set to 0.5 following \cite{belloni1}), the gravitational constant $G$, the mass $M_{\rm WD}$ and radius $R_{\rm WD}$ of the white dwarf, and the mass transfer rate $\dot{M}_{\rm dQ}$ from the donor star to the white dwarf during quiescence. The mass transfer rate $\dot{M}_{\rm A}$ establishes the limit between cold/neutral/stable and unstable discs:

\begin{equation}
\begin{aligned}
\dot{M}_{\rm A} = {} & 6.344 \times 10^{-11} \ \alpha_{\rm c}^{-0.004} \ \left( \frac{M_{\rm WD}}{\rm {\rm M_\odot}} \right)^{-0.88} \\
& \times \left( \frac{r}{10^{10} \ {\rm cm}} \right)^{2.65} \ {\rm M_\odot \ yr^{-1}}
\end{aligned}
\label{MA}
\end{equation}

Here, it is calculated using the alpha parameter $\alpha_{\rm c}$ related to the viscosity parameter in the accretion disc and adopted as 0.01, the mass $M_{\rm WD}$ of the white dwarf, and the distance $r$ from the centre of the white dwarf to the centre of mass. We also note that $\dot{M}_{\rm dQ}$ = $\dot{M}_{\rm A}$ is assumed for the simplification of accretion during the quiescence state.

We investigate the formation of GC CVs in two channels: 1. Primordially formed CVs, which develop from early binary systems inside the cluster, and 2. Dynamically formed/influenced CVs, which start from stellar interactions in the dense core such as binary exchanges and tidal captures. This distinction aids in understanding the impact of the cluster's dynamical environment on the CV population, differentiating between those evolving through isolated binary evolution and those resulting from the cluster’s internal dynamics.

\subsection{Simulated GC Analysis}

In this study, we used the MOCCA code to generate 81 simulated GC models and tracked their dynamics over a timescale of 12 Gyr, up to the PDP. The core density evolution across these models offers a key insight into the long-term dynamical behavior of these clusters. Specifically, we balanced the energy flow through the half-mass radius with the central energy generation from all sources. 

In \autoref{fig:model_division}, we show the temporal evolution of the core density $\rho_{c}$ in unit of $M_{\odot}$~pc$^{-3}$. According to their evolutions, we can divide our sample into three Classes: 

\begin{itemize}[leftmargin=*]
\item \textit{Class I:} Models with the core density decreases over time, implying expanding systems
\item \textit{Class II:} After expanding for a few Gyrs,  the core density starts increasing which implies the systems of this Class experience core collapsing. 
\item \textit{Class III:} 
Similar to Class II, but the systems of this Class have their core density reach a saturation point and start decreasing around the PDP, which indicates the core expansion is somehow resuscitated. 
\end{itemize}

Our classification respects the unique dynamical states at different stages of GC evolution. It acknowledges that Class II and Class III models undergo similar core density evolution but differ in the timing of energy depletion from core sources. In Class III, binaries become the main energy source again, particularly noticeable in the later stages of the model's evolution. Conversely, in the late phase of Class II, the core collapse generates potential energy, which marks a distinct difference from Class III, where the energy generation is primarily driven by binaries.

Mass segregation causes inward movement of massive objects which can impact the cluster's evolution significantly. The segregation of the most massive objects, namely the BH subsystems,  can suppress the core collapse of less massive systems and their dynamical interaction \citep{2022MNRAS.509.4713W}. Therefore, it is also important to investigate how the BH subsystems evolve in each Class.

In \autoref{fig:evolution_BH}, we trace the evolution of the BH population in Classes I/II/III. In this analysis, we account for every BH present in any system. The GCs of Class I are found to harbor the largest population of BH which supports the aforementioned notion that the presence of BH subsystems can result in the expansion of the lighter systems including CVs. On the other hand, BH population in Classes II/III is substantially smaller. Hence the core collapse of lighter systems can be triggered which results in the increasing $\rho_{c}$. In Class III, the GCs apparently turn from collapsing to expanding around the epoch of $\sim10-12$~Gyrs. This might suggest the binary burning of the low-mass binaries started driving the evolution of the GCs in this Class. This prompts us to investigate if there are any differences in the properties of CVs among these three Classes.

\begin{figure}
\includegraphics[width=\columnwidth]{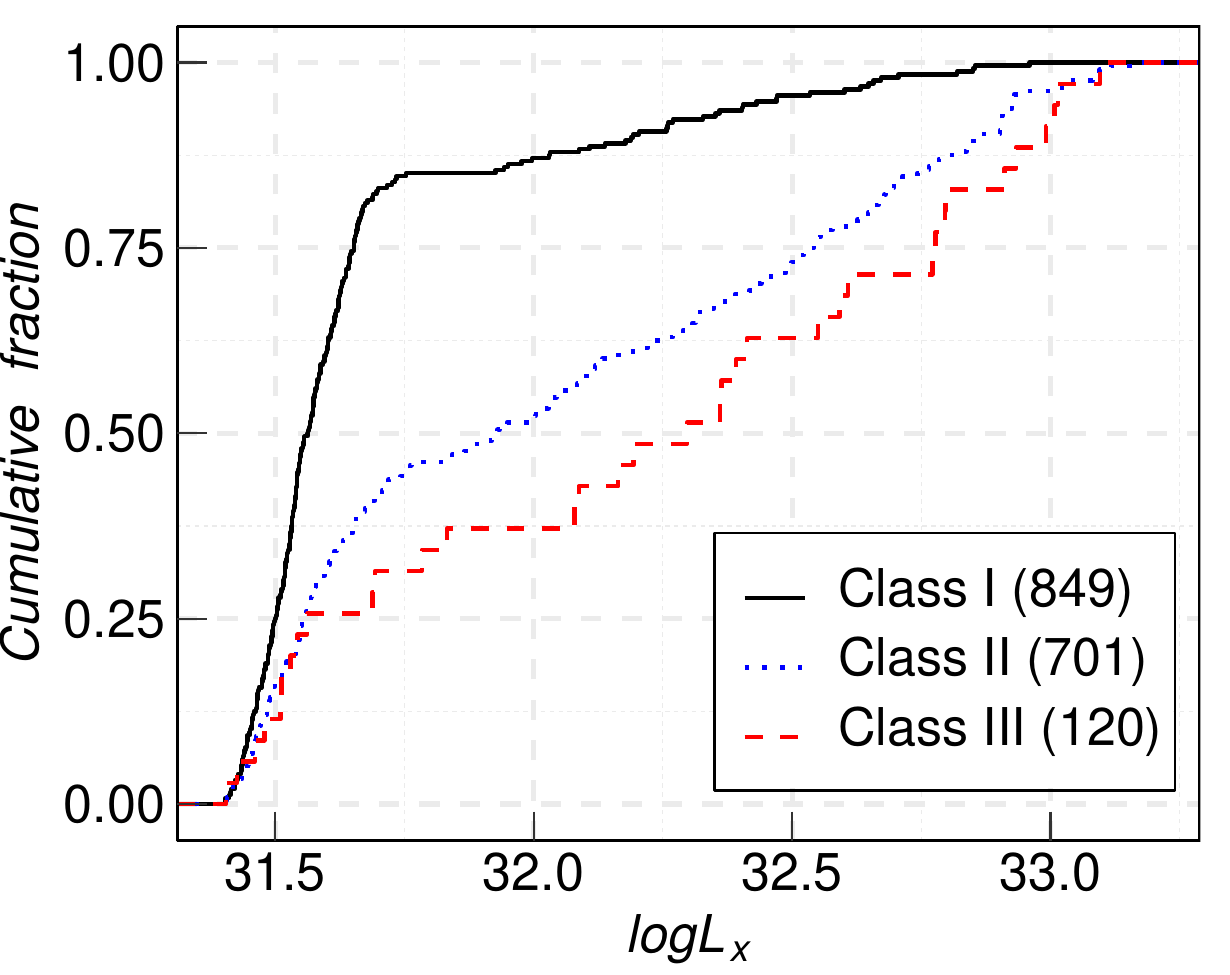}
\caption{eCDF of X-ray luminosity for CV populations of MOCCA across three classes defined by core density profiles at the PDP. It highlights differences among the groups, with Class III (the most dynamically evolved class) showcasing the most luminous CV population. This underscores the impact of dynamical evolution on CV properties. Bracketed numbers in the legends represent the sample sizes of CVs for each class.}
\label{fig:cdf_sim}
\end{figure}

In \autoref{fig:cdf_sim}, we present a plot to quantify the differences in the X-ray luminosity empirical distribution functions (eCDFs) of CV populations at the PDP. Among the three classes, Class III, which is considered the most dynamically evolved, displays the most luminous CV population. This emphasizes the influence of the dynamical state on the properties of CVs. To statistically validate this result, we employed a two-sample Anderson-Darling (A-D) test \citep{anderson_darling_1952}. This test was chosen for its sensitivity in detecting differences between eCDFs. We considered an A-D test $p$-value of less than 0.05 as evidence of a significant distinction between two eCDFs.

\begin{table}\centering
\begin{threeparttable}
\resizebox{0.97\columnwidth}{!}{
\begin{tabular}{ccc}
\toprule
Class I vs Class II\tnote{a} & Class I vs Class III\tnote{a} & Class II vs Class III\tnote{a}\\
\midrule
$5.8\times10^{-9}$  & $1.3\times10^{-5}$ & 0.44 \\
\bottomrule
\end{tabular}
}
\begin{tablenotes}
\item [a] {\scriptsize Three distinct populations of simulated GCs divided by core density evolution.}
\end{tablenotes}
\caption{Null hypothesis probabilities of the A-D test for comparing X-ray luminosity distributions among three different classes in MOCCA simulation data.}
\label{tab:adtest_sim}
\end{threeparttable}
\end{table}

Our results, which highlight the statistical differences between different populations, are summarized in \autoref{tab:adtest_sim}. These results indicate that Class III, which is generally considered to be the most dynamically evolved among the groups, has the most luminous CV population. This could be attributed to higher rates of interactions in their evolutionary pathways. In contrast, Class I shows significant differences when compared to both Class II and Class III, due to its less evolved dynamical state of the cluster, which results in a less luminous CV population. However, the $p$-value did not reveal a significant difference between Class II and Class III (i.e., $p > 0.05$).

\begin{figure}
\centering
\subfigure{
\includegraphics[width=0.95\columnwidth]{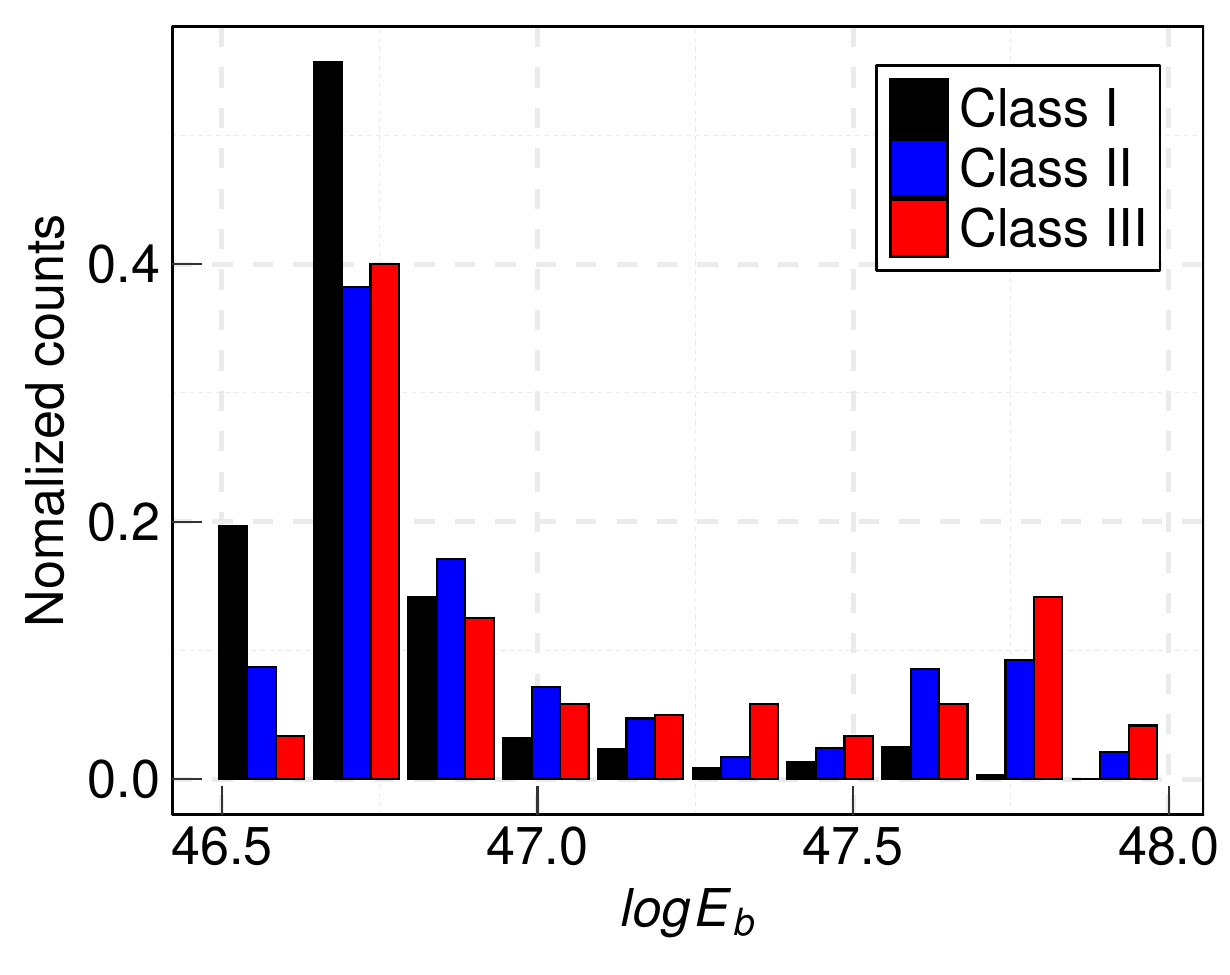}
}
\subfigure{
\includegraphics[width=0.95\columnwidth]{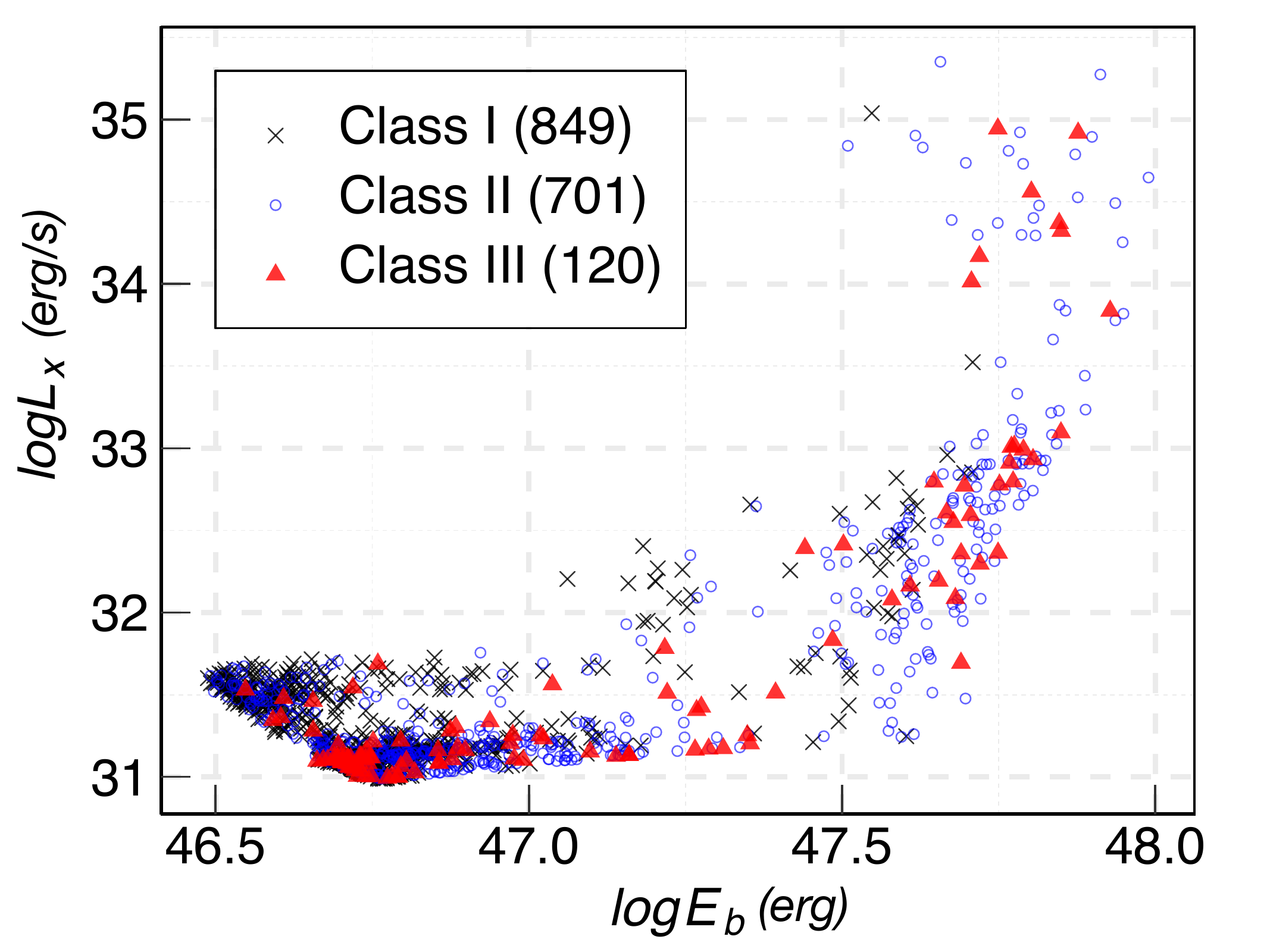}
}
\caption{Analysis of $E_b$ distributions in CVs at the PDP. The top panel depicts the bimodal $E_b$ distribution across CV classes, highlighting the absence of tightly-bound CVs in Class I. The lower panel reveals a correlation between higher binding energy and increased $L_x$ in CVs, indicating tighter orbits.}
\label{fig:binding}
\end{figure}

In the top panel of \autoref{fig:binding}, we show the binding energy $E_b$ distribution of CV populations for Classes I/II/III. It is clear that this distribution is bimodal with the second peak in the range of $E_{b}\sim10^{47.5}-10^{48.0}$~erg dominated by Classes II/III. This shows that such tightly-bounded CVs are absent in Class I, which suggests that their formation can be closely related to the dynamical status of their host GCs. The correlation between the $L_x$ and $E_{b}$ is shown in the lower panel of \autoref{fig:binding} highlights that the X-ray bright CVs are typically those with higher $E_{b}$, indicating tighter orbits.

\begin{figure}
\centering
\subfigure{
\includegraphics[width=0.95\columnwidth]{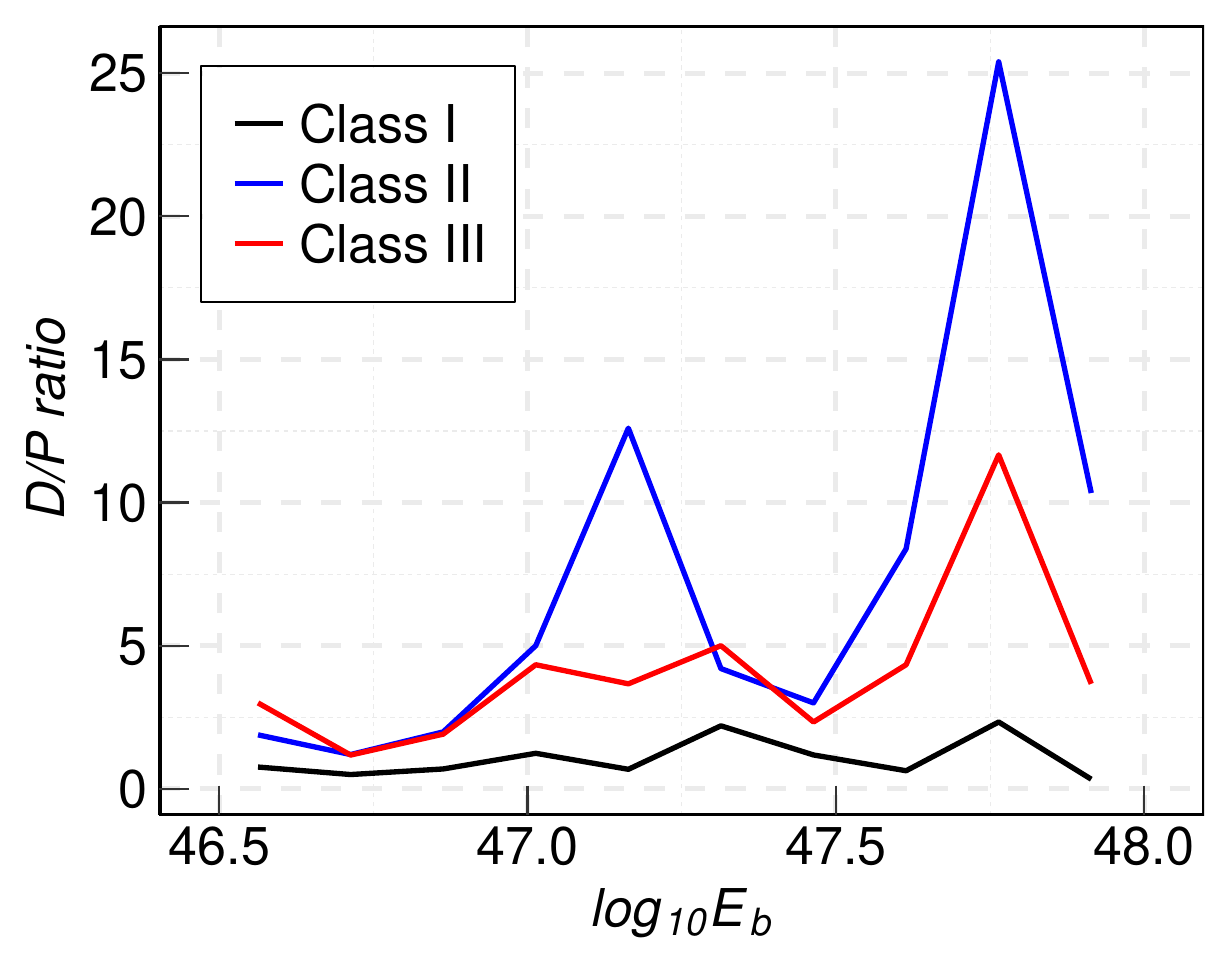}
}
\subfigure{
\includegraphics[width=0.95\columnwidth]{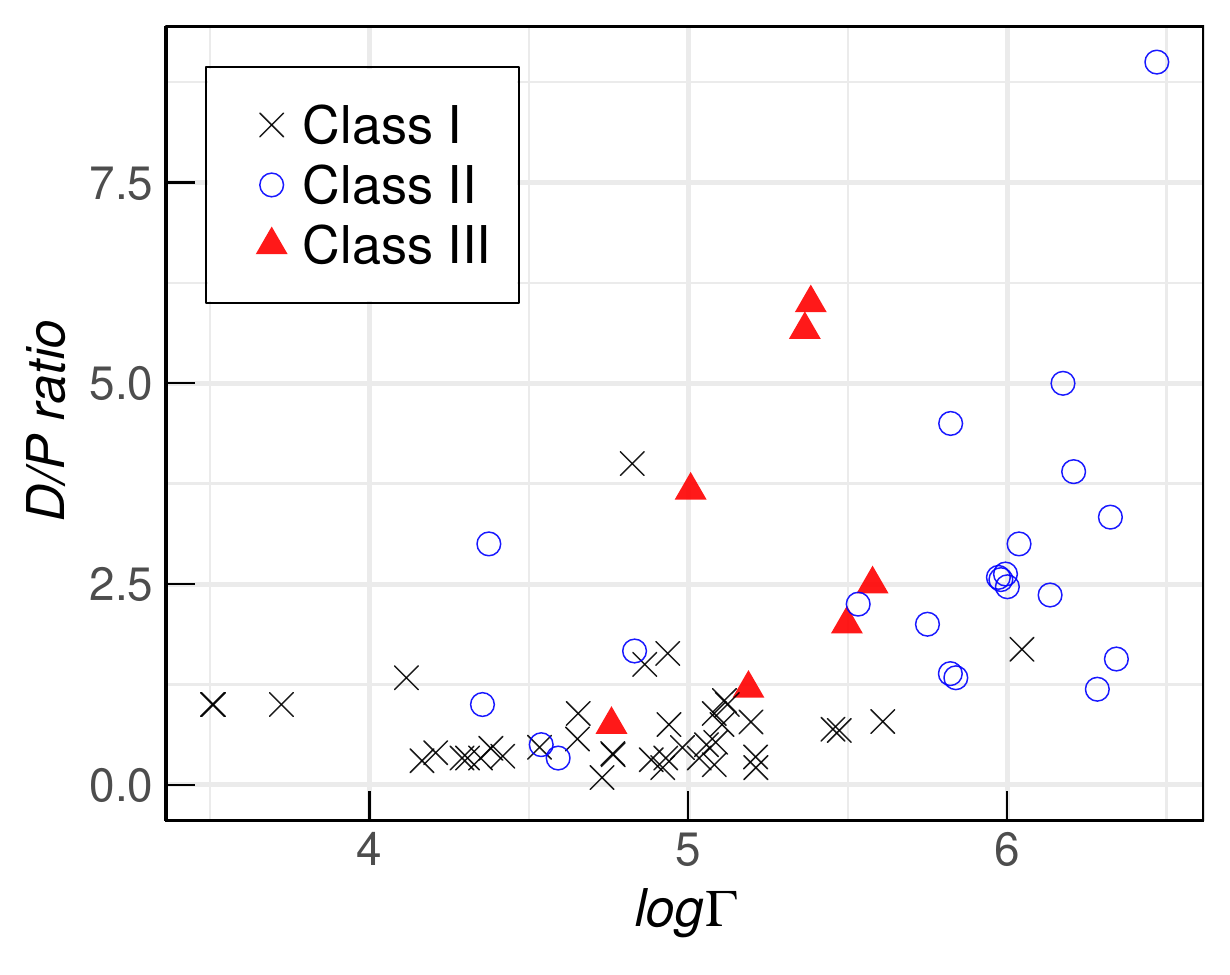}
}
\caption{The top panel shows the ratio of dynamically formed/perturbed CVs to the primordial CVs as a function of $E_b$ at the PDP. The lower panel presents this ratio as a function of the encounter rates ($\Gamma$), where $\Gamma \propto \int \rho^2/\sigma \, dr$ where $\rho$ is the central mass density and $\sigma$ is the velocity dispersion. Each point represents an individual cluster in each class at the PDP.}
\label{fig:dp_ratio}
\end{figure}

To further explore the dynamical effects on the formation of CVs, we investigate how the ratio between the dynamically formed/perturbed CVs and primordial CVs varies with $E_{b}$ in different Classes (See the top panel of \autoref{fig:dp_ratio}). It is evident that such a ratio for Class I is generally much lower than that of the other two Classes. This might suggest the dynamical formation of the CVs in Class I can be suppressed by the presence of BH subsystems. Additionally, this plot shows that the majority of the tightly-bounded CVs in Classes II/III are originated from the dynamical channels, which can be at least $\sim8$ times larger than their primordial counterparts in the same range of $E_{b}$. Presumably, the reduced number of BHs from the GCs in these two Classes might enable the core collapse of the light stars which subsequently facilitates the dynamical formation of low-mass binaries such as CVs \citep[cf.][]{2022MNRAS.509.4713W}. 

The lower panel of \autoref{fig:dp_ratio} shows the D/P ratio versus the encounter rates, with each point representing a cluster in different classes. The encounter rate ($\Gamma$) is determined by the integral of the stellar density squared divided by the velocity dispersion, specifically $\Gamma \propto \int \rho^2/\sigma, dr$, where $\rho$ is the stellar density and $\sigma$ is the velocity dispersion \citep{Bahramian_2013}. However, we used central mass density for $\rho$ in this study. For $\log \Gamma > 5.5$, the plot is dominated by Class II clusters, indicating they likely experience more stellar encounters. This increased encounter rate contributes to the higher D/P ratio observed in Class II clusters.

\section{X-ray Observations}
Apart from the simulations, we have also investigated the possible relationship between the dynamical state of GCs and the $L_{x}$ distribution of their hosted CVs with observational data. In order to resolve the X-ray point sources in GCs and analyze their $L_{x}$ and X-ray hardness, we have considered the data acquired by the Advanced CCD Imaging Spectrometer (ACIS) on board {\it Chandra}. The superior sub-arcsecond spatial resolution of ACIS enables the detection of point sources in the crowded environment of GCs. Furthermore, spectral information of the data allows us to construct the X-ray colour-luminosity distribution which we used for identifying the CV-like sources \citep[cf. ][]{Oh2020.498}. 

As we have mentioned in the introduction, based on the radial distribution of BSs, \cite{Ferraro_2012} suggests that GCs can be divided into three different dynamical-age families, which are referred to as Family I/II/III. Among 21 GCs in the sample used in their analysis, we found that 18 of them have been observed {\it Chandra} ACIS. The observation IDs of the data used in our work are summarized in \autoref{tab:datatable}. 

\begin{figure}
\includegraphics[width=0.9\columnwidth]{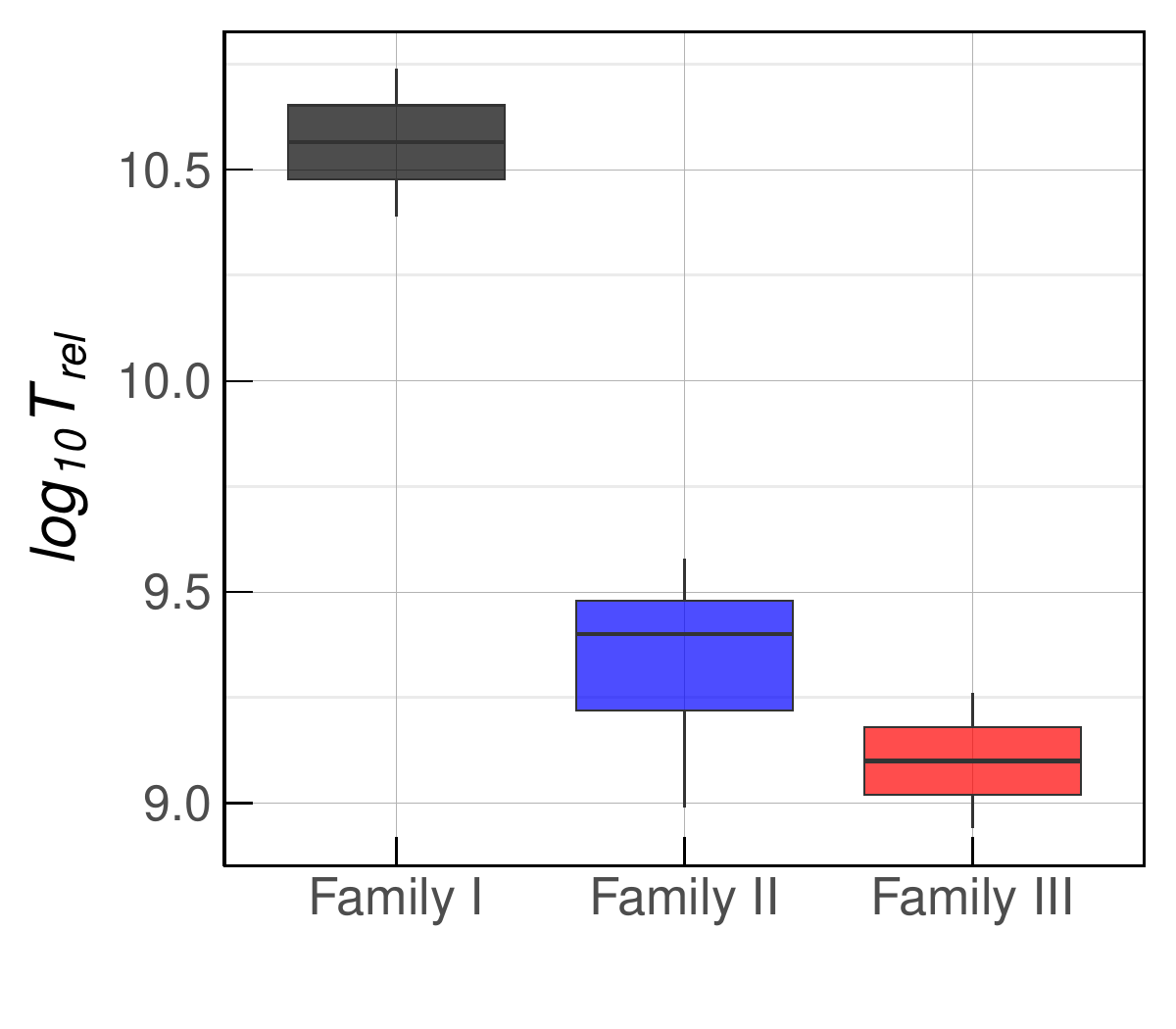}
\caption{Box plot of half-mass relaxation \citep{Baumgardt_2018} time between three different Families. Family I exhibits the longest half-mass relaxation time indicating a less dynamically evolved state.}
\label{fig:baumgardt}
\end{figure}

As a cross-check, we have also examined the half-mass relaxation time parameter from \cite{Baumgardt_2018} to test whether the Family division was well-separated. \autoref{fig:baumgardt} shows the half-mass relaxation times ($T_{rel}$) across the three families. We can see that $T_{rel}$ of Family I is much longer than the other two Families which indicates it is less dynamically evolved. And hence, this plot supports the division scheme adopted by \cite{Ferraro_2012}.

\subsection{Data Reduction \& Analysis}

\begin{table}\centering
\begin{threeparttable}[b]
\begin{tabular}[t]{clp{30 mm}c}
\toprule\toprule
ID          & Name & ObsID                        & Family \\ 
\midrule\midrule
\multicolumn{4}{c}{GCs in \cite{Ferraro_2012}} \\ 
\midrule
$\omega$-Cen  & NGC 5139$^{\rm a}$ & 1519, 13726, 13727           & I      \\ 
\midrule
  -           & NGC 2419           & 10490                        & I      \\ 
\midrule
47 Tuc        & NGC 104$^{\rm b}$  & 78, 953, 954, 955, 966, 2735, 2736, 2737, 2738, 3384, 3385, 3386, 3387, 15747, 15748, 16527, 16528, 16529, 17420, 26229, 26286 & II     \\ 
\midrule
  -           & NGC 6752$^{\rm c}$ & 948, 6612, 19013, 19014, 20121, 20122, 20123                     & II      \\ 
\midrule
M 3           & NGC 5272$^{\rm d}$ & 4542, 4543, 4544             & II      \\ 
\midrule
M 4           & NGC 6121$^{\rm e}$ & 946, 7446, 7447              & II      \\ 
\midrule
M 13          & NGC 6205           & 5436, 7290                   & II      \\ 
\midrule
M 92          & NGC 6341$^{\rm l}$           & 3778, 5241                   & II      \\ 
\midrule
  -           & NGC 6388$^{\rm f}$ & 5505, 12453                  & II      \\ 
\midrule
  -           & NGC 288$^{\rm g}$  & 3777                         & II      \\ 
\midrule
M 53          & NGC 5024           & 6560                         & II      \\ 
\midrule
M 5           & NGC 5904$^{\rm d}$ & 2676                         & II      \\ 
\midrule
M 10          & NGC 6254           & 16714                        & II      \\ 
\midrule
M 55          & NGC 6809$^{\rm h}$ & 4531                         & II      \\ 
\midrule
M 2           & NGC 7089           & 8960                         & II      \\
\midrule
M 30          & NGC 7099$^{\rm i}$ & 2679, 18997, 20725, 20726, 20731, 20732, 20792, 20795, 20796     & III    \\ 
\midrule
M 79          & NGC 1904           & 9027                         & III     \\ 
\midrule
M 80          & NGC 6093           & 1007                         & III     \\
\bottomrule\bottomrule
\end{tabular}
    \vspace{2mm}
    \footnotesize{$^{\rm a}$ \cite{10.1093/mnras/sty675},}
    \footnotesize{$^{\rm b}$ \cite{Heinke_2005},\cite{RiveraSandoval:2017itj},}
    \footnotesize{$^{\rm c}$ \cite{10.1093/mnras/stu559},}
    \footnotesize{$^{\rm d}$ \cite{Tweddale},}
    \footnotesize{$^{\rm e}$ \cite{Bassa_2004},}
    \footnotesize{$^{\rm l}$ \cite{Lu_2011}.}
    \footnotesize{$^{\rm f}$ \cite{Maxwell_2012},}
    \footnotesize{$^{\rm g}$ \cite{Kong_2006},}
    \footnotesize{$^{\rm h}$ \cite{Bassa_2008},}
    \footnotesize{$^{\rm i}$ \cite{Lugger_2007}.}
    \caption{ObsIDs of Chandra X-ray Observatory for a total 18 GCs in this study. The data set forms the foundation for the analysis of X-ray luminosities and the investigation of CV populations in relation to their host GCs.}
    \label{tab:datatable}
\end{threeparttable}
\end{table}

For data processing and analysis, we utilized the {\it Chandra} Interactive Analysis of Observation software (CIAO version 4.15.1) \citep{CIAO_2006} with the recently updated calibration database\footnote{\url{https://cxc.cfa.harvard.edu/caldb/downloads/release_notes/CALDB_v4.10.4.html}}. All the data were reprocessed by using the CIAO script \texttt{chandra\_repro}.
Since most X-ray sources are located within the half-mass radius $r_{h}$, we have selected the circular regions with the angular size corresponding to $r_{h}$ in each data before proceeding to further analysis. We have further restricted all the subsequent analyses to an energy range of 0.3-7 keV. 

In case there are multiple observations of a given GC, we have created the X-ray image by using the CIAO tool \texttt{merge\_obs} tool with \texttt{binsize=0.5} for attaining subpixel resolution. For each image of GC, 
we have run the wavelet detection algorithm (CIAO tool \texttt{wavdetect}) on all the images with a range of scales (1, 1.414, 2, 2.828). We have adopted a significance threshold of $10^{-6}$ which corresponds to $\sim1$ false alarm for running the algorithm on an image with $1k\times1k$ pixels.
We considered an X-ray source to be genuine if it is detected at a significance larger than $3\sigma$. This procedure results in the detection of 334 X-ray sources from 18 GCs in our sample.

We utilized the CIAO script \texttt{srcflux} to calculate the unabsorbed net X-ray fluxes $f_{x}$ of these sources in 0.3-7~keV. For absorption correction, we have estimated the column absorption $n_{H}$ towards each GC by the relation $n_{H}=6.86\times10^{21}E(B-V)$~cm$^{-2}$ \citep[cf.][]{NH_redden} where $E(B-V)$ is the foreground reddening as reported by \cite{Harris2010arXiv1012.3224H}. For all sources residing in the same GC, we assume they have the same $n_{H}$. To estimate $f_{x}$, we assume an absorbed power-law model with $n_{H}$ fixed by the aforementioned procedure and the photon index $\Gamma$ fixed at 1.7 \citep[cf.][]{Oh2020.498}. For background subtraction, we have adopted the source-free regions around each GC. In order to compute the X-ray colour, we have divided the absorption-corrected X-ray fluxes into the soft band (0.3-1.5~keV) $f_{x}^{\rm S}$ and hard band (1.5-7.0 keV) $f_{x}^{\rm H}$.
The X-ray luminosity is calculated by $L_{x}=4\pi d^{2}f_{x}$, where $d$ is the distance of the GC \citep{Harris2010arXiv1012.3224H} from the solar system. On the other hand, the X-ray colour is defined as $2.5\log\left(f_{x}^{\rm S}/f_{x}^{\rm H}\right)$.

Using the X-ray colours and $L_{x}$ obtained from the aforementioned procedures, we construct the X-ray colour-luminosity diagram. This enables us to select promising CV candidates as different classes of X-ray binaries are found to occupy different regions in the colour-luminosity diagram \cite[cf.][and references therein]{Oh2020.498}. Searching for the relevant literature, we found that 55 X-ray sources in our sample have been classified as CV through multi-wavelength analysis. The properties of these X-ray emitting GC CVs are summarized in \autoref{tab:CVs}.

In order to expand the sample size for a constraining statistical analysis, we have attempted to select more CV-like X-ray sources through the techniques of machine learning. For building up the training/test samples for supervised classification, besides the aforementioned 55 confirmed CVs, we have also collected the data from 59 non-CV X-ray sources with confirmed nature from those 18 GCs under our consideration. These non-CV X-ray sources include MSPs, LMXBs, and active binaries (ABs). 70\% of these samples are used for training. And the rest are taken as a test set.  

\begin{figure*}
\centering
\subfigure{
\includegraphics[width=0.18\textwidth]{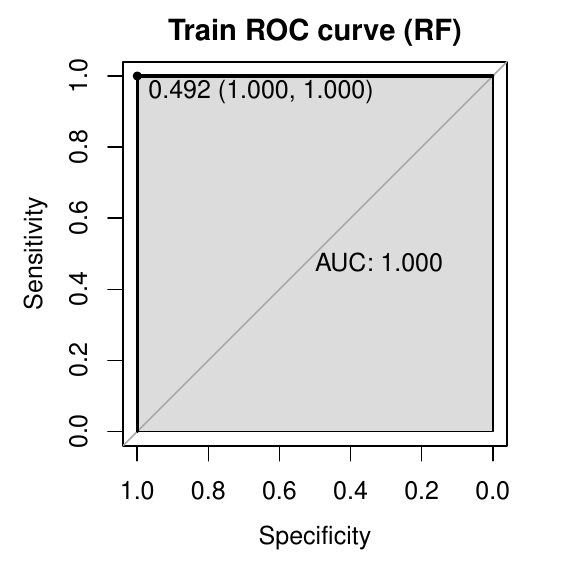}
}
\subfigure{
\includegraphics[width=0.18\textwidth]{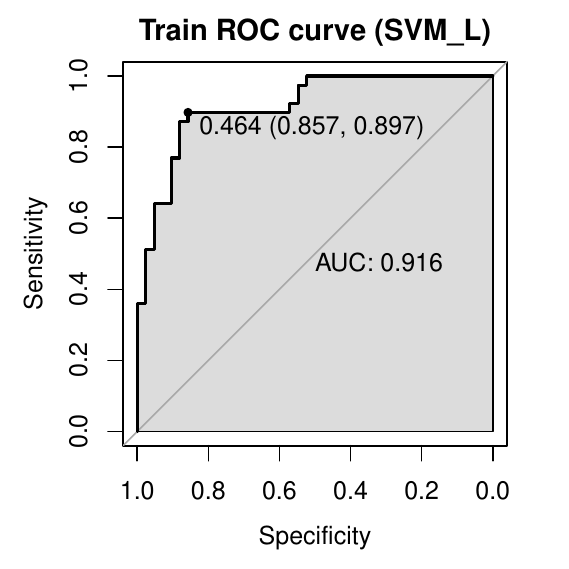}
}
\subfigure{
\includegraphics[width=0.18\textwidth]{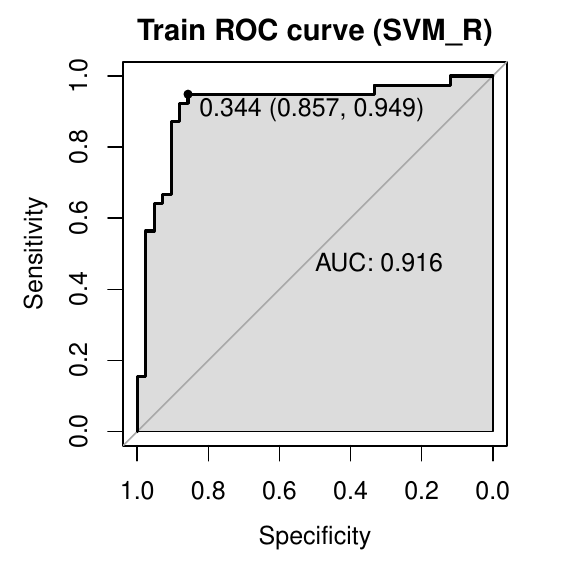}
}
\subfigure{
\includegraphics[width=0.18\textwidth]{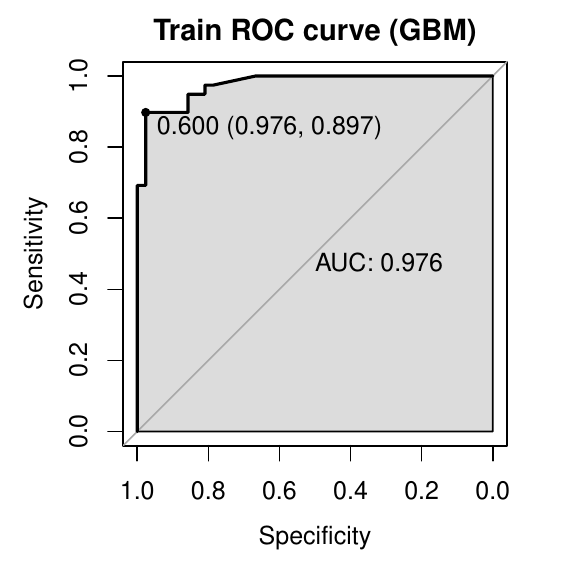}
}
\subfigure{
\includegraphics[width=0.18\textwidth]{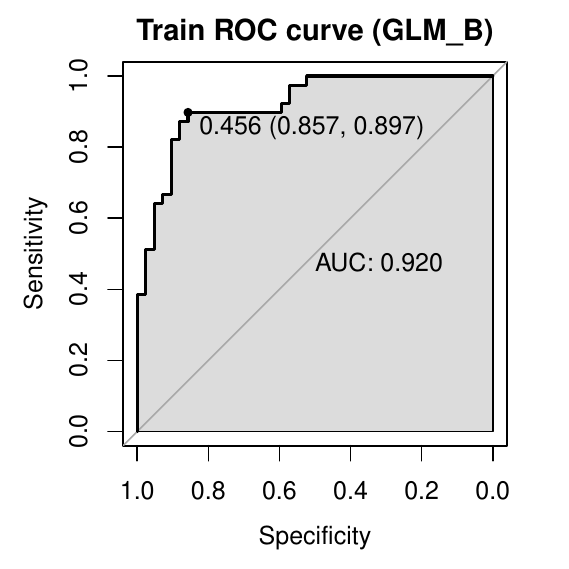}
}
\subfigure{
\includegraphics[width=0.18\textwidth]{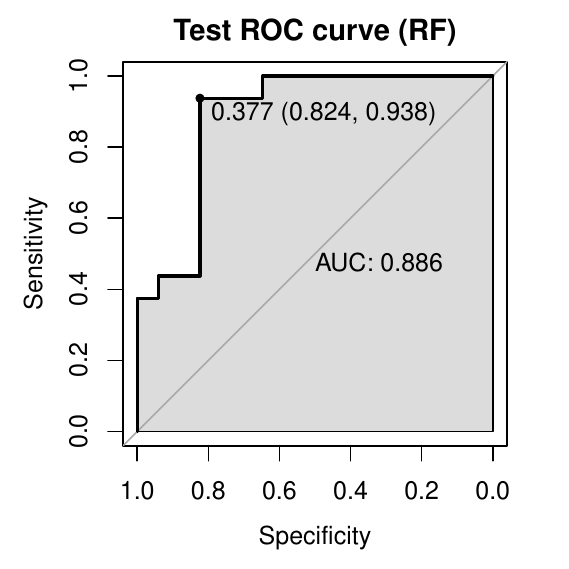}
}
\subfigure{
\includegraphics[width=0.18\textwidth]{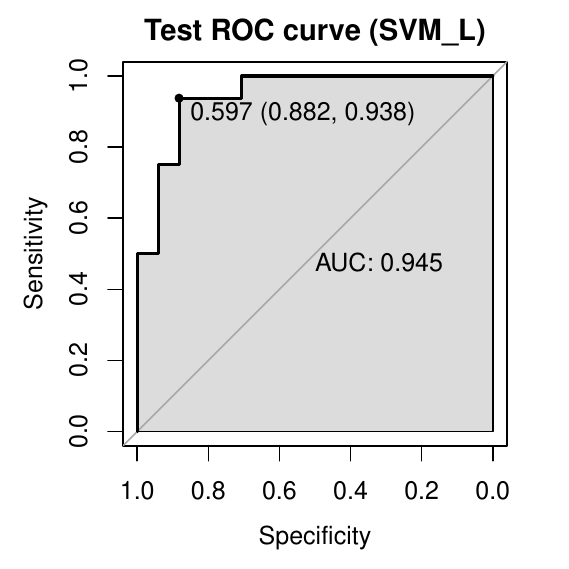}
}
\subfigure{
\includegraphics[width=0.18\textwidth]{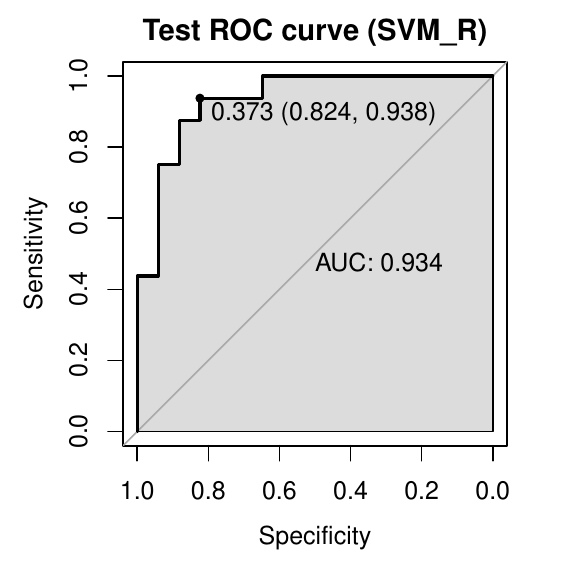}
}
\subfigure{
\includegraphics[width=0.18\textwidth]{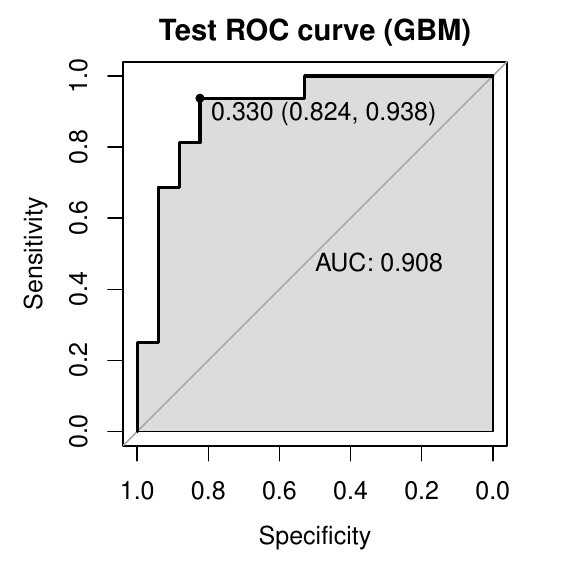}
}
\subfigure{
\includegraphics[width=0.18\textwidth]{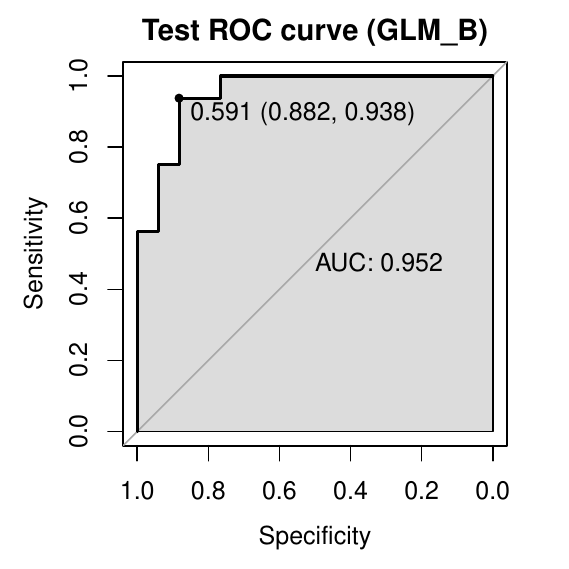}
}
\caption{The training and test ROC curves for five different classifiers in our ensemble for the CV/Non-CV classification.}
\label{fig:ML}
\end{figure*}

With a view of obtaining a more robust predictive performance, instead of relying on a single classifier, we adopted the approach of ensemble learning by combining the predictions from multiple models. Our ensemble contains five different classifiers: random forest (RF), supporting vector machine (SVM) with linear kernel, SVM with radial kernel, gradient boosting machine (GBM), and boosted generalized linear model (GLMBoost). To determine the optimal threshold of each classifier, we adopted the procedure of Leave-One-Out Cross-Validation (LOOCV). In \autoref{fig:ML}, we show the Receiver Operating Characteristic (ROC) curves for each classifier with both training set and test set. ROC curve is a plot of sensitivity (i.e. true positive rate) against specificity (i.e. one minus false positive rate). The Area Under the Curve (AUC) of an ROC curve provides an aggregated performance measure for all possible thresholds. An ideal classifier is expected to yield an AUC close to unity. The test ROC of all classifiers in our ensemble attains the AUC in a range of 0.89-0.95. We also quantify the performance of each classifier with the overall accuracy, which is defined to be the percentage of correct classification in the test set. The resultant accuracy of our selected classifiers ranges from 88\% to 91\%. All these performance metrics suggest that the CV/Non-CV classification is reasonable. 

\begin{figure}
\includegraphics[width=\columnwidth]{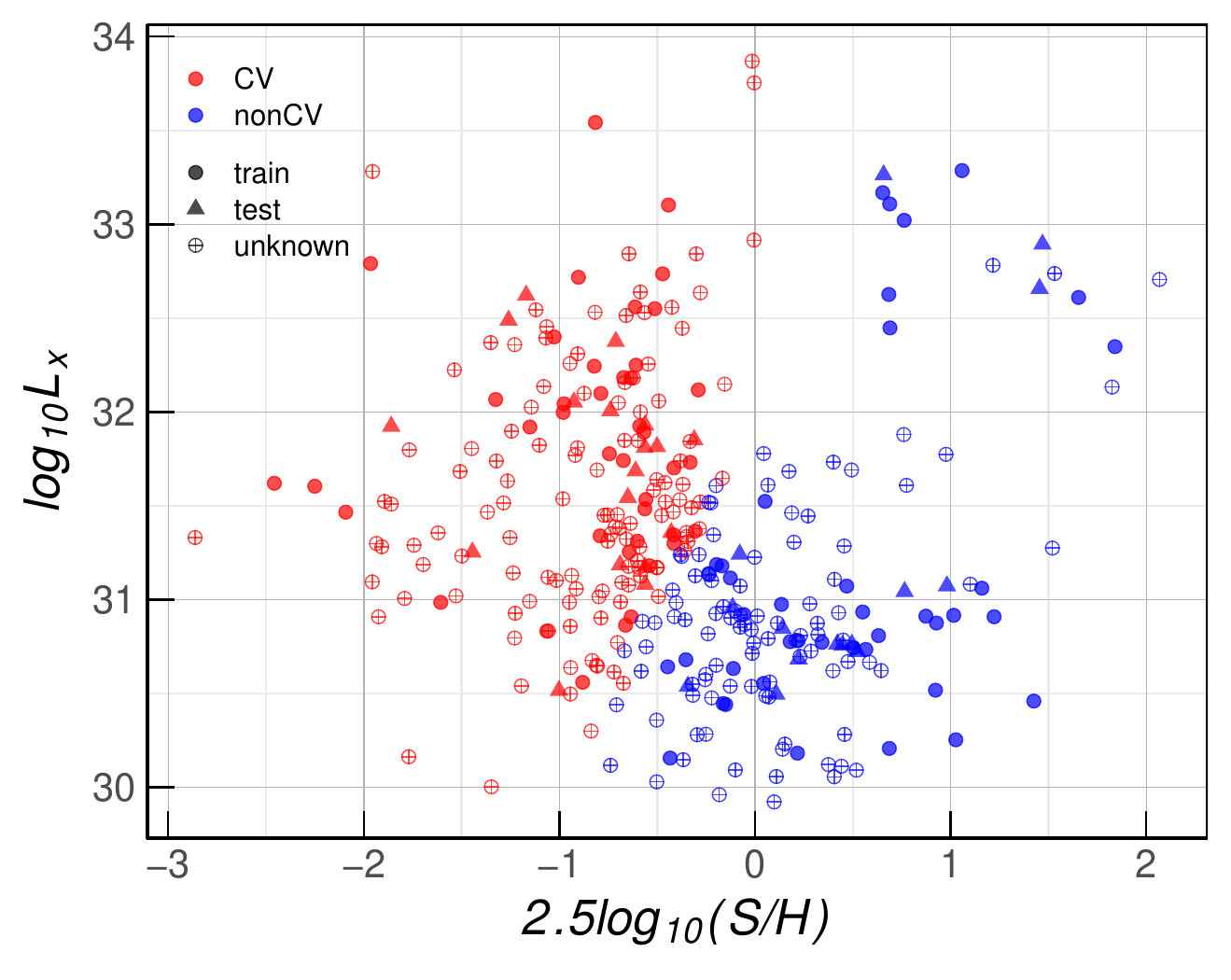}
\caption{The X-ray colour-luminosity diagram of 334 X-ray sources detected in 18 GCs. 55 confirmed CVs are illustrated by the red circles/triangles and the data in training/test sets. On the other hand, 59 Non-CVs with confirmed nature are illustrated by the blue circles/triangles and the data in training/test sets. The other 220 unidentified X-rays are illustrated by the open symbols. With our trained ensemble, these sources are labeled as CV-like/Non-CV-like with red/blue colour.}
\label{fig:CMD_binary}
\end{figure}

For a robust prediction on the unidentified X-ray sources, we adopted the procedure of majority voting for aggregating the predictions from different classifiers into an ensemble model. The ensemble model gives an overall accuracy of 88\%. The classification of the training set, test set, as well as unidentified X-ray sources obtained by the ensemble model, is shown in \autoref{fig:CMD_binary}. By applying this model to select CV candidates from the unidentified X-ray sources, we have obtained an additional 124 CV-like sources. To enlarge the sample size of our statistical analysis, we append these CV candidates to our sample.

\begin{figure}
\includegraphics[width=\columnwidth]{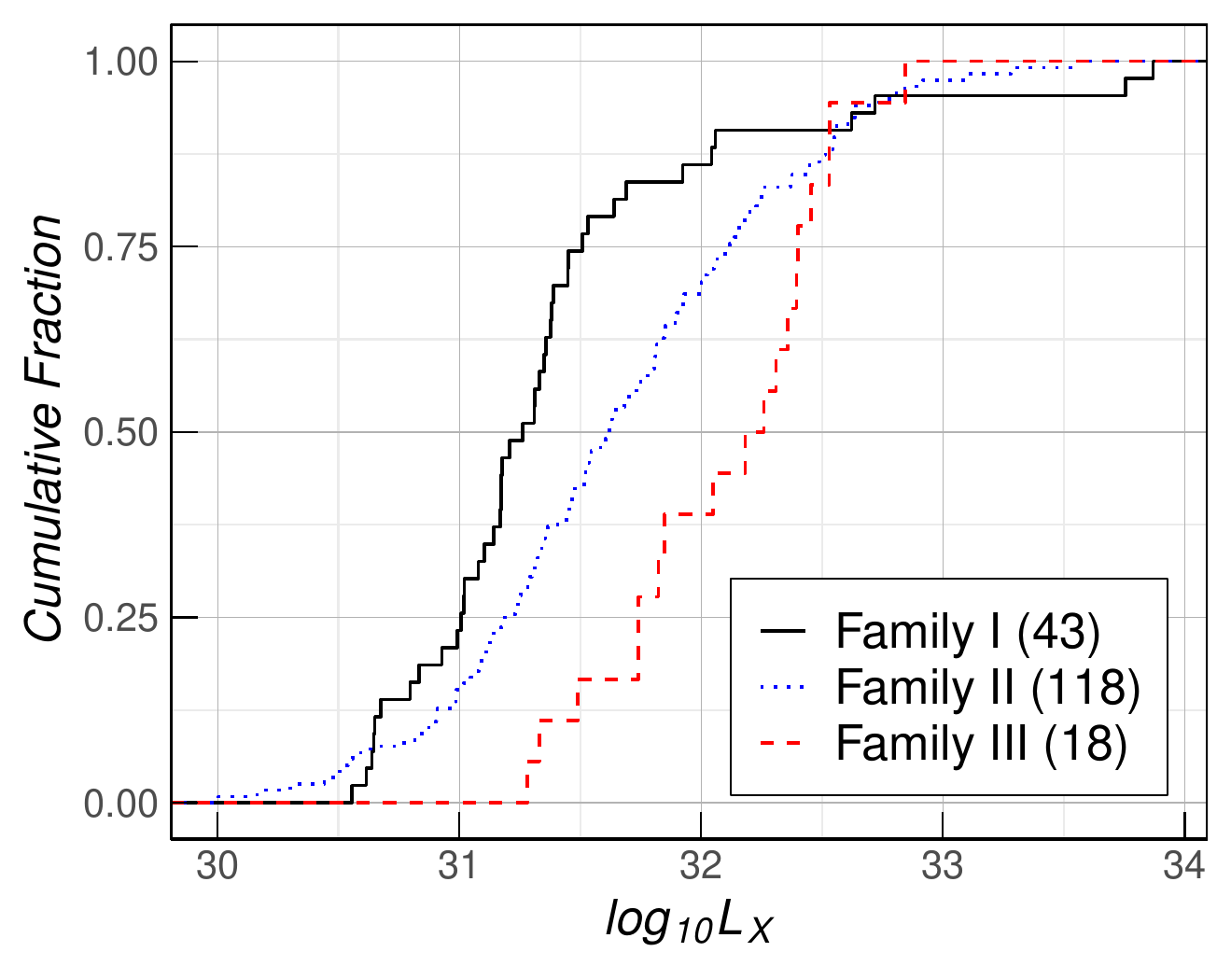}
\caption{
Observed eCDF of X-ray luminosity for the three different populations of CV candidates in GCs. The dynamically older population (Family III) exhibits the brightest CV candidates among the three families, highlighting the impact of the dynamical state on the properties of CVs. The bracketed numbers in the legends show the corresponding sample sizes of CV candidates.}
\label{fig:cdf_obs}
\end{figure}

Dividing this enlarged sample into three groups of different dynamical ages according to \cite{Ferraro_2012}, we have 43, 118, and 18 CV-like X-ray sources for Families I, II, and III, respectively. We proceeded to examine whether there is any significant difference in the distribution of $L_{x}$ among these three Families. The visual comparisons of the eCDFs of $L_{x}$ among Family I/II/III are shown in \autoref{fig:cdf_obs}. Apparently, the dynamically older population (Family III) hosts more luminous CVs. To quantify the significance of such differences, we have performed two-sample A-D tests. The $p-$values yielded by such test are summarized in \autoref{tab:adtest_obs}. We found that the $L_{x}$ distribution is significantly different among these three Families. Apart from the results as yielded by the sample selected by the ensemble model, we also tabulate the $p-$values obtained by the samples selected by the individual classifier in \autoref{tab:adtest_obs} which all show consistent results.

\section{Summary \& Discussion}
This study embarked on a comprehensive investigation into the relationship between the dynamical state of GCs and the properties of their CV populations. Our approach has been two-pronged: we have conducted both simulations and observational studies to determine if these two investigating directions can lead to a consistent picture. In this section, we summarized our results and discussed the possible implications.

We utilized the MOCCA code for our simulations, providing us with a robust tool for simulating star cluster evolution, incorporating both stellar dynamics and stellar evolution processes. These simulations enabled us to model the evolution of GCs and their CV populations through the X-ray luminosity, offering a theoretical framework for understanding the complex interplay between GC dynamics and the characteristics of their constituent stars, especially CVs. We divided the simulated GCs into three different populations according to their core density evolution.

Our simulation results have demonstrated that the dynamical evolution of GCs has a profound impact on the properties of CVs. We found significant differences between Class I and both Class II and Class III, indicating different evolutionary rates and suggesting a correlation with the core density evolution. However, there was no notable differences were found between Class II and Class III. In \autoref{fig:binding}, the bottom panel shows that the bright population of CVs has a tighter orbit than those of a less luminous population. It is important to note that when considering the number in each class, CVs in Class III would exhibit a larger bright population than those in Class II.

On the other hand, our analysis of {\it Chandra} data has involved the examination of 179 CV candidates in 18 GCs as selected by supervised machine learning. By dividing them into three families corresponding to different dynamical ages \cite{Ferraro_2012}, we have revealed a clear connection between the dynamical state of GCs and the X-ray luminosity distributions of their CVs, with the most dynamically evolved GCs displaying the most X-ray luminous CV population.

Our observational investigation has revealed a correlation between the dynamical age of a GC and the X-ray luminosity distribution of CV candidates. While \cite{Ferraro_2012} derive Families I/II/III based on the properties of BSs, we have found another class of binaries can also exhibit different properties among these three families. Assuming such a correlation can be generalized, this might suggest that one can infer the dynamical status of a GC from its binary population. For example, one can draw insight into the dynamical age of a GC by comparing the X-ray eCDF of CVs to the distributions given in \autoref{fig:cdf_obs}. 

Despite that our results follow the classification given by \cite{Ferraro_2012}, we have to highlight one possible caveat of this scheme. \cite{Ferraro_2012} divide the GCs into Families I/II/III according to the spatial distribution of BSs. \cite{2017MNRAS.471.2537H} have attempted to reproduce this result with both MOCCA and N-body simulations. However, they suggested that the bimodal spatial distribution of BSs pointed out by \cite{Ferraro_2012} can be a transient feature. This can hamper the robustness of the classification scheme of \cite{Ferraro_2012}. 
Although the X-ray luminosity distributions of CVs also show the difference among Families I/II/III, further investigations are still required to verify this scheme. Apart from a single parameter (i.e. $L_{x}$), analyses with other observable properties from the compact binaries are encouraged to examine the validity of the classification scheme given by \cite{Ferraro_2012}. 

\begin{table}\centering
\begin{threeparttable}
\resizebox{0.97\columnwidth}{!}{
\begin{tabular}{c|ccc}
\toprule
& Family I vs Family II\tnote{a} & Family I vs Family III\tnote{a} & Family II vs Family III\tnote{a}\\
\midrule
{Ensemble} & $1.9\times10^{-3}$ & $1.2\times10^{-5}$ & $5.4\times10^{-3}$ \\
RF & $3.1\times10^{-4}$ & $6.7\times10^{-7}$ & $3.9\times10^{-3}$ \\
SVM (linear) & $1.0\times10^{-2}$ & $6.6\times10^{-5}$ & $4.3\times10^{-3}$ \\
SVM (radial) & $4.2\times10^{-3}$ & $1.3\times10^{-5}$ & $2.5\times10^{-3}$ \\
GBM & $1.7\times10^{-2}$ & $8.0\times10^{-6}$ & $1.1\times10^{-3}$ \\
GLB-Boost & $1.3\times10^{-2}$ & $9.2\times10^{-5}$ & $4.8\times10^{-3}$ \\
\bottomrule
\end{tabular}
}
\begin{tablenotes}
\item [a] {\scriptsize Dynamical-age family defined by \protect\cite{Ferraro_2012}.}
\end{tablenotes}
\caption{Null hypothesis probabilities of the A-D test for comparing X-ray luminosity distributions among three different families. The results include not only the robust classification of CV from the ensemble model but also comparative outcomes from individual classifiers.}
\label{tab:adtest_obs}
\end{threeparttable}
\end{table}

While the results inferred from both simulation and observational data suggest the X-ray luminosity distribution of CVs in a GC is closely related to its dynamical state, one cannot unambiguously establish a one-to-one correspondence between the simulation and the actual X-ray observation. This limitation is due to the complexities in fully capturing the co-evolution of X-ray binaries and intracluster dynamics in simulations. First, the eCDFs for Classes I/II/III in \autoref{fig:cdf_sim} do not match with those for Families I/II/III in \autoref{fig:cdf_obs}. This can be ascribed to the fact that our current simulation cannot fully capture the complexity of the co-evolution between the X-ray binaries and the intracluster dynamics. Although our simulations provide valuable perspectives on the dynamical intricacies of GCs and their CV populations, they are based on an earlier version of the BSE code which does not fully incorporate the roles of BHs in the mass segregation. With advancements in simulation fidelity and a deeper understanding of dynamical processes, further research using updated simulation codes is essential to enhance our understanding of the co-evolution of GCs and their X-ray binaries.
 
Additionally, during the pre-main-sequence (PMS) phase, complexities increase due to the larger size of stellar objects, affecting tightly orbiting or highly eccentric binaries. This stage, known as PMS eigen-evolution \citep{Kroupa_1995}, involves mass transfer from high mass star to low mass companion and orbit circularization, impacting the binary's evolutionary path, which can hinder CV development in isolated binary systems \citep{belloni1,Hong_2017}. The dynamical nature of dense environments thus plays a crucial role in creating conducive conditions for CV formation, underscoring the importance of these interactions in our simulations. Nevertheless, the role of the initial binary distribution remains a subject of debate, highlighting the need for continued research in this area.

\begin{table*}
\centering
\begin{threeparttable}
\renewcommand{\arraystretch}{1.3}
\begin{tabularx}{\textwidth}{>{\centering\arraybackslash}p{1cm} >{\centering\arraybackslash}X >{\centering\arraybackslash}X >{\centering\arraybackslash}X >{\centering\arraybackslash}X>{\centering\arraybackslash}X>{\centering\arraybackslash}X}
\toprule\toprule
GC & RA & Dec & R/${R_c}$ & $L_x$ & Source significance & Reference \\ 
\midrule\midrule

47Tuc	&00:24:3.8754	&-73:53:38.2608	 & 4.14	    &31.35	$^{+0.04}_{-0.03}$	&52.9   &  ${\rm a}$  \\
47Tuc	&00:24:11.1218	&-73:53:39.8860	 & 4.22	    &31.25	$^{+0.04}_{-0.04}$	&49.23  &  ${\rm a}$  \\
47Tuc	&00:24:7.1610	&-73:54:14.0992	 & 2.49 	&31.82	$^{+0.02}_{-0.02}$	&117.78 &  ${\rm a}$  \\
47Tuc	&00:24:9.2416	&-73:54:15.9927	 & 2.50 	&30.91	$^{+0.05}_{-0.05}$	&26.89  &  ${\rm a}$  \\
47Tuc	&00:24:2.1432	&-73:54:17.8615	 & 2.41 	&32.24	$^{+0.01}_{-0.01}$	&210.15 &  ${\rm a}$  \\
47Tuc	&00:24:7.7858	&-73:54:32.6349	 & 1.67 	&31.24	$^{+0.04}_{-0.03}$	&47.33  &  ${\rm a}$  \\
47Tuc	&00:24:2.5504	&-73:54:48.5854	 & 1.10 	&31.54	$^{+0.02}_{-0.02}$	&96.13  &  ${\rm a}$  \\
47Tuc	&00:24:8.4908	&-73:54:59.4846	 & 0.71 	&31.47	$^{+0.03}_{-0.02}$	&81.98  &  ${\rm a}$  \\
47Tuc	&00:24:3.7009	&-73:55:0.8549	 & 0.52  	&31.08	$^{+0.04}_{-0.04}$	&11.62  &  ${\rm a}$  \\
47Tuc	&00:24:4.2679	&-73:55:1.7844	 & 0.40	    &33.10	$^{+0.01}_{-0.01}$	&577.24 &  ${\rm a}$  \\
47Tuc	&00:24:6.0181	&-73:55:3.6042	 & 0.19	    &31.93	$^{+0.01}_{-0.02}$	&134.04 &  ${\rm a}$  \\
47Tuc	&00:24:4.9216	&-73:55:4.3243	 & 0.21	    &31.62	$^{+0.02}_{-0.02}$	&55.74  &  ${\rm a}$  \\
47Tuc	&00:24:2.8283	&-73:55:10.7176	 & 0.63	    &31.85	$^{+0.02}_{-0.01}$	&96.6   &  ${\rm a}$  \\
47Tuc	&00:24:3.1389	&-73:55:12.4948	 & 0.59	    &30.56	$^{+0.07}_{-0.08}$	&9.25   &  ${\rm a}$  \\
47Tuc	&00:24:6.3973	&-73:55:16.8434	 & 0.46	    &32.45	$^{+0.01}_{-0.01}$	&266.85 &  ${\rm a}$  \\
47Tuc	&00:24:7.8105	&-73:55:18.2622	 & 0.68	    &31.74	$^{+0.02}_{-0.02}$	&108.23 &  ${\rm a}$  \\
47Tuc	&00:24:15.8987	&-73:55:23.4478	 & 2.31	    &32.00	$^{+0.01}_{-0.02}$	&163.75 &  ${\rm a}$  \\
47Tuc	&00:24:8.3097	&-73:55:24.0148	 & 0.95	    &31.31	$^{+0.04}_{-0.02}$	&48.2   &  ${\rm a}$  \\
47Tuc	&00:24:16.9929	&-73:55:32.6883	 & 2.69	    &31.93	$^{+0.02}_{-0.01}$	&157.65 &  ${\rm a}$  \\
47Tuc	&00:24:10.7630	&-73:55:34.1166	 & 1.65	    &31.60	$^{+0.02}_{-0.02}$	&103.16 &  ${\rm a}$  \\
47Tuc	&00:24:3.7807	&-73:55:36.8891	 & 1.42	    &31.89	$^{+0.02}_{-0.01}$	&144.73 &  ${\rm a}$  \\
47Tuc	&00:24:5.3945	&-73:55:38.2562	 & 1.43	    &30.99	$^{+0.05}_{-0.04}$	&33.71  &  ${\rm a}$  \\
47Tuc	&00:24:4.1586	&-73:54:47.2874	 & 0.99	    &30.44	$^{+0.08}_{-0.08}$	&7.54   &  ${\rm a}$  \\
47Tuc	&00:24:5.0097	&-73:54:53.5541	 & 0.66	    &30.63	$^{+0.08}_{-0.06}$	&13.35  &  ${\rm a}$  \\
47Tuc	&00:24:5.0030	&-73:56:29.9067	 & 3.82	    &30.52	$^{+0.12}_{-0.11}$	&7.4    &  ${\rm a}$  \\
M3  	&13:42:9.7751	&+28:22:47.7569	 & 1.18	    &33.54	$^{+0.02}_{-0.02}$	&118.92 &  ${\rm b}$  \\
M30 	&21:40:22.1775	&-24:49:7.7821	 & 1.33	    &32.18	$^{+0.03}_{-0.03}$	&61.16  &  ${\rm c}$  \\
M30 	&21:40:22.9568	&-24:49:10.2117	 & 3.27	    &32.40	$^{+0.03}_{-0.02}$	&91.01  &  ${\rm c}$  \\
M4  	&16:23:34.1325	&-27:28:25.2475	 & 0.21	    &31.92	$^{+0.01}_{-0.01}$	&140.18 &  ${\rm d}$  \\
M4  	&16:23:46.4080	&-27:28:44.4438	 & 2.17	    &31.37	$^{+0.04}_{-0.03}$	&35.26  &  ${\rm d}$  \\
M4  	&16:23:34.3247	&-27:29:20.8002	 & 0.79	    &31.18	$^{+0.04}_{-0.04}$	&38.92  &  ${\rm d}$  \\
M5  	&15:18:34.4323	&+02:05:6.8599	 & 0.90	    &32.07	$^{+0.08}_{-0.06}$	&21.18  &  ${\rm b}$  \\
M55 	&19:40:8.6156	&-31:01:7.8777	 & 1.19	    &32.01	$^{+0.07}_{-0.07}$	&21.25  &  ${\rm e}$  \\
M92 	&17:17:6.4931	&+43:08:3.4147	 & 0.74	    &31.70	$^{+0.13}_{-0.11}$	&14.29  &  ${\rm f}$  \\
M92 	&17:17:7.2967	&+43:08:7.1378	 & 0.16	    &32.25	$^{+0.06}_{-0.06}$	&31.15  &  ${\rm f}$  \\
M92 	&17:17:7.1447	&+43:08:13.1197	 & 0.29	    &32.63	$^{+0.04}_{-0.03}$	&52.18  &  ${\rm f}$  \\
M92 	&17:17:6.8991	&+43:08:35.0671	 & 1.68	    &31.25	$^{+0.23}_{-0.13}$	&7.02   &  ${\rm f}$  \\
NGC288	&00:52:44.6887	&-27:25:11.6735	 & 0.14	    &32.49	$^{+0.06}_{-0.04}$	&47.43  &  ${\rm g}$  \\
NGC5139	&13:26:52.1358	&-48:30:24.5517	 & 0.49	    &32.72	$^{+0.01}_{-0.01}$	&180.78 &  ${\rm h}$  \\
NGC5139	&13:26:53.5147	&-48:30:59.8014	 & 0.46	    &32.62	$^{+0.01}_{-0.01}$	&173.48 &  ${\rm h}$  \\
NGC5139	&13:26:28.6561	&-48:33:32.8858	 & 1.65	    &31.92	$^{+0.03}_{-0.03}$	&36.38  &  ${\rm h}$  \\
NGC5139	&13:26:38.4285	&-48:29:23.5619	 & 1.00	    &31.17	$^{+0.09}_{-0.06}$	&17.61  &  ${\rm h}$  \\
NGC5139	&13:26:20.3769	&-48:29:57.1984	 & 1.99	    &32.04	$^{+0.02}_{-0.02}$	&28.28  &  ${\rm h}$  \\
NGC6388	&17:36:16.6384	&-45:15:36.5946	 & 2.34	    &32.79	$^{+0.04}_{-0.05}$	&35.64  &  ${\rm i}$   \\
NGC6388	&17:36:15.7578	&-45:15:44.9582	 & 2.40	    &32.74	$^{+0.04}_{-0.05}$	&39.05  &  ${\rm i}$   \\
NGC6388	&17:36:17.2540	&-45:15:49.6535	 & 0.36	    &32.56	$^{+0.09}_{-0.06}$	&10.43  &  ${\rm i}$   \\
NGC6388	&17:36:17.1640	&-45:15:58.2700	 & 0.85	    &32.55	$^{+0.08}_{-0.05}$	&15.19  &  ${\rm i}$   \\
NGC6752	&19:10:56.0147	&-60:00:22.6628	 & 4.32	    &32.18	$^{+0.02}_{-0.01}$	&163.91 &  ${\rm j}$  \\
NGC6752	&19:10:51.5128	&-60:00:32.8886	 & 2.27	    &31.73	$^{+0.02}_{-0.02}$	&87.97  &  ${\rm j}$  \\
NGC6752	&19:10:54.7765	&-60:00:46.1302	 & 2.17	    &31.34	$^{+0.04}_{-0.04}$	&51.16  &  ${\rm j}$  \\
\bottomrule\bottomrule
\end{tabularx}
\vspace{2mm}
\end{threeparttable}
\end{table*}

\begin{table*}
\centering
\begin{threeparttable}
\renewcommand{\arraystretch}{1.3}
\begin{tabularx}{\textwidth}{>{\centering\arraybackslash}p{1cm} >{\centering\arraybackslash}X >{\centering\arraybackslash}X >{\centering\arraybackslash}X >{\centering\arraybackslash}X>{\centering\arraybackslash}X>{\centering\arraybackslash}X}
\toprule\toprule
GC & RA & Dec & R/${R_c}$ & $L_x$ & Source significance & Reference \\ 
\midrule\midrule
NGC6752	&19:10:51.1506	&-60:00:48.0482	 & 1.02	    &32.38	$^{+0.01}_{-0.01}$	&196.62  &  ${\rm j}$  \\
NGC6752	&19:10:51.5982	&-60:00:58.2446	 & 0.46	    &32.05	$^{+0.02}_{-0.02}$	&120.7   &  ${\rm j}$  \\
NGC6752	&19:10:51.5205	&-60:01:3.1266	 & 0.86	    &31.81	$^{+0.02}_{-0.02}$	&92.32   &  ${\rm j}$  \\
NGC6752	&19:10:40.3884	&-60:01:18.5590	 & 8.91	    &32.12	$^{+0.02}_{-0.02}$	&130.12  &  ${\rm j}$  \\
NGC6752	&19:10:40.6348	&-61:59:53.9478	 & 10.38	&30.83	$^{+0.09}_{-0.06}$	&18.66   &  ${\rm j}$  \\
\bottomrule\bottomrule
\end{tabularx}
\vspace{2mm}
    \footnotesize{CV references: $^{\rm a}$ \cite{Heinke_2005},\cite{RiveraSandoval:2017itj},}
    \footnotesize{$^{\rm b}$ \cite{Tweddale},}
    \footnotesize{$^{\rm c}$ \cite{Lugger_2007},}
    \footnotesize{$^{\rm d}$ \cite{Bassa_2004},}
    \footnotesize{$^{\rm e}$ \cite{Bassa_2008},}
    \footnotesize{$^{\rm f}$ \cite{Lu_2011},} 
    \footnotesize{$^{\rm g}$ \cite{Kong_2006},} 
    \footnotesize{$^{\rm h}$ \cite{10.1093/mnras/sty675},}
    \footnotesize{$^{\rm i}$ \cite{Maxwell_2012},} 
    \footnotesize{$^{\rm j}$ \cite{10.1093/mnras/stu559}.}
    \\
    \footnotesize{Note: RA and Dec are given in J2000 coordinates. R/${R_c}$ represents the radial distance from the cluster center normalized by the core radius in parsec. $L_x$ is the logarithmically scaled scaled X-ray luminosity in units of ${\text{erg/s}}$. Source significance is given in sigma ($\sigma$) levels.}\
\caption{Confirmed CVs in Milky Way Globular clusters}
\label{tab:CVs}
\end{threeparttable}
\end{table*}

\section*{Acknowledgements}

K.OH is supported by the National Research Foundation of Korea grant 2022R1A6A3A13071461. C.Y.H. is supported by the research fund of Chungnam National University and by the National Research Foundation of Korea grant 2022R1F1A1073952. MIG is supported by the grant 2021/41/B/ST9/01191 financed by the Polish National Science Centre (NCN).

\section*{Data Availability}
The data underlying this article were accessed from Chandra Data Archive (https://cda.harvard.edu/chaser/).






\bsp	
\label{lastpage}
\end{document}